\documentclass{article}%
\usepackage{amsmath}
\usepackage{amsmath,extarrows}
\usepackage{amsfonts}
\usepackage{amssymb}
\usepackage{amsthm}
\usepackage{mathdots}
\usepackage{graphicx}
\usepackage[normalem]{ulem}
\usepackage{color}
\usepackage{hyperref}
\usepackage{setspace}
\usepackage{array}
\usepackage{graphicx}
\usepackage{multirow}
\usepackage{tikz}
\usepackage{cases}
\usepackage{enumitem}
\usepackage{fullpage}%
\providecommand{\U}[1]{\protect\rule{.1in}{.1in}}
\usetikzlibrary{decorations.pathreplacing}
\usetikzlibrary{positioning, shapes.geometric}
\tikzset{global scale/.style={
scale=#1,
every node/.append style={scale=#1}}}
\providecommand{\U}[1]{\protect\rule{.1in}{.1in}}
\allowdisplaybreaks
\newtheorem{theorem}{Theorem}

\newtheorem{corollary}[theorem]{Corollary}

\newtheorem{definition}[theorem]{Definition}
\newtheorem{example}[theorem]{Example}

\newtheorem{lemma}[theorem]{Lemma}
\newtheorem{notation}[theorem]{Notation}

\newtheorem{proposition}[theorem]{Proposition}
\newtheorem{remark}[theorem]{Remark}

\newtheorem{assumption}[theorem]{Assumption}

\begin{document}

\title{A Generalization of Habicht's Theorem \\for Subresultants of Several Univariate Polynomials}
\author{Hoon Hong\\Department of Mathematics, North Carolina State University\\Box 8205, Raleigh, NC 27695, USA\\hong@ncsu.edu\\[10pt] Jiaqi Meng\thanks{Corresponding author.}, Jing Yang\\SMS--HCIC--School of Mathematics and Physics,\\Center for Applied Mathematics of Guangxi,\\Guangxi Minzu University, Nanning 530006, China\\mjq252920@163.com; yangjing0930@gmail.com\\[-10pt]}
\date{}
\maketitle

\begin{abstract}
Subresultants of two univariate polynomials are one of the most classic and
ubiquitous objects in computational algebra and algebraic geometry. In 1948,
Habicht discovered and proved interesting relationships among subresultants.
Those relationships were found to be useful for both structural understanding
and efficient computation.

\

Often one needs to consider several (possibly more than two) polynomials. It
is rather straightforward to generalize the notion of subresultants to several
polynomials. However, it is not obvious (in fact, quite challenging) to
generalize the Habicht's result to several polynomials. The main contribution
of this paper is to provide such a generalization.

\end{abstract}


\section{Introduction}

Subresultants of two polynomials are one of the most ubiquitous objects in
computational algebra and algebraic geometry. It played a vital role in the
development of many fundamental algorithms, such as triangular decomposition,
quantifier elimination and parametric gcd. Therefore, extensive studies have
been carried out on the underlying theories, efficient algorithms and various
applications (just list a few
~\cite{Barnett:1971,Bostan:2017,Brown:1971,Buse:2004a,Collins:1967,Cox:2023,DAndrea:2001,DAndrea:2007,DAndrea:2009,DAndrea:2006,DAndrea:2013,DAndrea:2015,DAndrea:2019,Ho:1989,Hong:1996a,Hong:1999a,Hong:2001a,Hong:2001b,Hong_Yang-2023,Householder:1968,Kakie:1976,Sylvester:1853,Vardulakis:1978,Wang:1998,Wang:2000,Wang:2024}%
,
\cite{Bostan2020,Buse:2004b,Hou:2000,Jaroschek:2013,Lascoux:2003,Roy:2020,Terui:2008}
and \cite{Diaz-Toca:2004,Krick:2017,Li:1998}).

In 1948, Habicht discovered and proved interesting relationships among
subresultants, which is the well known Habicht's theorem. The classical
Habicht's theorem contains two results:

\begin{itemize}
\item relationship between a single subresultant and the pseudo-remainder of
its two consecutive subresultants ;

\item relationship between a single subresultant and the subresultant of two others.
\end{itemize}

\noindent In \cite{Hong_Yang-2023}, Hong and Yang generalize the
\textit{first} relationship to subresultants of \textit{several} polynomials.
In the current paper, we will generalize the \textit{second} one to
\textit{several} polynomials.

\bigskip

To set a suitable context, let us review Habicht's second relation. Let
$F_{0},F_{1}$ be two univariate polynomials of degrees $d_{0}\leq d_{1}$
without losing generality. Let $R_{k}\left(  F_{0},F_{1}\right)  $ denote the
$\left(  d_{0}-k\right)  $-th subresultant of~$F_{0}$and~$F_{1}$. (Note that
we re-index the subresultant because it is helpful in generalizing the concept
to several polynomials). Habicht posed the following interesting question: are
there integers $u,v,w_{0},w_{1}$ and a constant $c$ such that the following
equality holds?
\begin{equation}
c\;R_{u}(F_{0},F_{1})\;\;=\;\;R_{v}(R_{w_{0}}(F_{0},F_{1}),R_{w_{1}}%
(F_{0},F_{1})) \label{eqs:relation_2polys}%
\end{equation}
If so, what are conditions on them? Habicht gave the following elegant answer
in \cite{Habicht:1948}: If
\[
\left\{
\begin{array}
[c]{rcl}%
w_{1} & = & w_{0}+1\\
v & \geq & 1\\
u & = & w_{0}+v
\end{array}
\right.
\]
then the equality \eqref{eqs:relation_2polys} holds where $c$ is some power of
the principal coefficient of $R_{w_{0}}(F_{0},F_{1})$ in terms of~$x$. This
inherent relationship says that a subresultant of lower degree can be computed
from those of higher degree. By applying the relation repeatedly, one can find
more relations.

In the case of more than two polynomials, one could compute the subresultants
in a recursive way. However, it results in nested subresultants and often
causes an exponential expansion of degree in parameters. Thus one is keen on
constructing subresultants in a non-recursive way. For this purpose, one needs
to define the notion of subresultant for several polynomials. It can be done
in a very natural way (see \cite{Hong_Yang-2023}).

Let $R_{k}$ denote for the $k$-th subresultant of several polynomials (of
course, $k$ is no longer a single number, but could be a vector of numbers).
We pose the following question/challenge: are there integer vectors
$u,v,w_{0},\ldots,w_{n}$ and a constant $c$ such that the following holds?
\begin{equation}
c\;R_{u}(F_{0},\ldots,F_{n})\;\;=\;\;R_{v}(R_{w_{0}}(F_{0},\ldots
,F_{n}),\ldots,R_{w_{n}}(F_{0},\ldots,F_{n})) \label{eqs:relation_npolys}%
\end{equation}
If so, what are conditions on them? In this paper, we give a (hopefully)
elegant answer: If%
\[
\left\{
\begin{array}
[c]{rcl}%
w_{1} & = & w_{0}+e_{1}\\
& \vdots & \\
w_{n} & = & w_{0}+e_{n}\\
v & = & (k,\ldots,k)+e_{i}\\
u & = & w_{0}+v
\end{array}
\right.
\]
for some $k\geq1$ and $0\leq i\leq n$ , then the equality
\eqref{eqs:relation_npolys} holds where $c$ is some power of the principal
coefficient of $R_{w_{0}}(F_{0},\ldots,F_{n})$ in terms of $x$ and where
$e_{j}\ $stands\ for\ the\ $j$-th\ unit\ vector\ of\ length\ $n$.

Note that the generalized Habicht's theorem (given in this paper) has the
essentially same structure as the original Habicht's theorem, i.e., a
subresultant of several polynomials can be computed from some of their
subresultants with higher degree of the given polynomials. By applying the
relation repeatedly, one can find more relations.

The main difficulty for tackling this problem is that we have to overcome the
following challenge. In the case of two polynomials, in order to identify the
conditions for \eqref{eqs:relation_2polys}, one needs to consider nested
subresultants, i.e., subresultants of subresultants, which are very
complicated in their form. Thus, the multi-polynomial case faces the same kind
of difficulty but in a much larger scale.

\medskip\noindent\textsf{Related works:}

\begin{enumerate}
\item In \cite{Hong_Yang-2023}, for computing the greatest common divisor of
several univariate polynomials with coefficients in an efficient way, Hong and
Yang identified a relationship between subresultants and pseudo-remainders of
these subresultants for several polynomials. Since the pseudo-remainder of two
polynomials can be viewed as a special form of their subresultants, the
relationship in \cite{Hong_Yang-2023} can be viewed as a special case of the
equality \eqref{eqs:relation_npolys}. In this paper, we give a much more
general condition for \eqref{eqs:relation_npolys} to hold.

\item In \cite{1995_Chardin, 2001_DAndrea_Dickenstein, 2006_Andrea_Krick_Szanto, 1990_Gonzalez_Vega, 2008_Szanto, 2010_Szanto} and \cite{Buse:2004a,Cox:2023}, the authors generalized the subresultant of two
univariate polynomials to that of  multivariate polynomials while
constraining the number of polynomials to be at most one more than the number
of variables and investigated their algebraic properties such as irreducibility and relationship
to a shape lemma.  In this paper, we take another route for generalization. We generalize
the subresultant of two
univariate polynomials to several  polynomials while staying univariate and investigate
their underlying structure in order to generalize Habicht's theorem to several univariate
polynomials.
\end{enumerate}

\noindent The paper is structured as follows. In Section \ref{sec:preliminaries}, we
review some concepts of subresultants for several univariate polynomials. Then
the main result is presented in Section \ref{sec:main_result}. The following
Section \ref{sec:remarks} provides a thorough explanation of its relationship
with the classical Habicht's theorem. The proof of the main result is given in
Section \ref{sec:proof}. The paper is concluded in Section
\ref{sec:conclusion}.

\section{Preliminaries}

\label{sec:preliminaries} In this section, we review the followings: (1)
subresultants of two polynomials and Habicht's theorem on them, and (2) a
natural generalization of subresultants to several polynomials. We will do so
using a \textbf{new indexing} scheme for them. The reason is that the new
indexing will facilitate the generalization to several polynomials. Thus, we
strongly encourage the readers (even though who know the classical theory)
read this section in order to get familiar with the new indexing scheme.

Let $\mathcal{Z}$ denote an integral domain such as $\mathbb{Z}$, $\mathbb{Q}%
$, $\mathbb{Z}\left[  a\right]  $ and so on. The followings are taken directly
from~\cite{Hong_Yang-2023}. For readers' convenience, we reproduce them here.


\subsection{Review on subresultants of two polynomials}

\begin{definition}
[Determinant polynomial of matrix]Let $M\in\mathcal{Z}^{p\times q}$ where
$p\leq q$ (that is, $M$ is square or wide).

\begin{itemize}
\item The \emph{determinant polynomial} of $M$, written as
${\operatorname*{dp}}(M)$, is defined by
\[
{\operatorname*{dp}}(M)=\sum_{0\leq j\leq q-p}c_{j}x^{j}%
\]
where $c_{j}=\det\left[  M_{1}\ \cdots\ M_{p-1}\ M_{q-j}\right]  $ and $M_{k}$
stands for the $k$-th column of $M$.

\item The \emph{principal coefficient} of $\operatorname*{dp}(M)$, written as
$\operatorname*{pcdp}(M)$, is defined by
\[
{\operatorname*{pcdp}}\left(  M\right)  =\ \operatorname*{coeff}%
\nolimits_{x^{q-p} }\left(  \operatorname*{dp}\left(  M\right)  \right)
\]

\end{itemize}
\end{definition}

\begin{example}
Let%
\[
M=\left[
\begin{array}
[c]{ccccc}%
m_{11} & m_{12} & m_{13} & m_{14} & m_{15}\\
m_{21} & m_{22} & m_{23} & m_{24} & m_{25}\\
m_{31} & m_{32} & m_{33} & m_{34} & m_{35}%
\end{array}
\right]
\]
Note $p=3$ and $q=5$. Thus

\begin{itemize}
\item ${\operatorname*{dp}}\left(  M\right)  =c_{2}x^{2}+c_{1}x^{1}+c_{0}
x^{0}$ where
\begin{align*}
c_{2}  &  =\det\left[  M_{1}\ M_{2}\ M_{5-2}\right]  =\det\left[
\begin{array}
[c]{cc|c}%
m_{11} & m_{12} & m_{13}\\
m_{21} & m_{22} & m_{23}\\
m_{31} & m_{32} & m_{33}%
\end{array}
\right] \\
c_{1}  &  =\det\left[  M_{1}\ M_{2}\ M_{5-1}\right]  =\det\left[
\begin{array}
[c]{cc|c}%
m_{11} & m_{12} & m_{14}\\
m_{21} & m_{22} & m_{24}\\
m_{31} & m_{32} & m_{34}%
\end{array}
\right] \\
c_{0}  &  =\det\left[  M_{1}\ M_{2}\ M_{5-0}\right]  =\det\left[
\begin{array}
[c]{cc|c}%
m_{11} & m_{12} & m_{15}\\
m_{21} & m_{22} & m_{25}\\
m_{31} & m_{32} & m_{35}%
\end{array}
\right]
\end{align*}

\item ${\operatorname*{pcdp}}\left(  M\right)  =c_{2}=\det\left[  M_{1}%
\ M_{2}\ M_{5-2}\right]  =\det\left[
\begin{array}
[c]{cc|c}%
m_{11} & m_{12} & m_{13}\\
m_{21} & m_{22} & m_{23}\\
m_{31} & m_{32} & m_{33}%
\end{array}
\right]  $
\end{itemize}
\end{example}

\begin{definition}
[Coefficient matrix of a list of polynomials]Let $P=(P_{1},\ldots,P_{t})$
where
\[
P_{i}=\sum_{0\le j\le p_{i}}b_{ij}x^{j}\in\mathcal{Z}[x]
\]
and $p_{i}=\deg P_{i}$. Let $m=\max_{1\le i\le t} p_{i}$. Then the
\emph{coefficient matrix} of $P$, written as ${\operatorname*{cm}}(P)$, is
defined as the $t\times(m+1)$ matrix whose $(i,j)$-th entry is the coefficient
of $P_{i}$ in the term $x^{m+1-j}$.
\end{definition}

\begin{example}
\label{example:P-cm(P)} Let $P=(P_{1},P_{2},P_{3})$ where
\begin{align*}
P_{1}  &  =b_{03}x^{3}+b_{02}x^{2}+b_{01}x+b_{00}\\
P_{2}  &  =b_{13}x^{3}+b_{12}x^{2}+b_{11}x+b_{10}\\
P_{3}  &  =b_{22}x^{2}+b_{21}x+b_{20}%
\end{align*}
Thus
\begin{align*}
\operatorname{cm}(P)=\operatorname{cm}(P_{1},P_{2},P_{3})= \left[
\begin{array}
[c]{cccc}%
b_{03} & b_{02} & b_{01} & b_{00}\\
b_{13} & b_{12} & b_{11} & b_{10}\\
& b_{22} & b_{21} & b_{20}%
\end{array}
\right]
\end{align*}

\end{example}

\begin{notation}
[Determinant polynomial of a list of polynomials]Let $P=(P_{1},\ldots,P_{t})$
be such that $\operatorname*{cm}(P)$ is square or wide. Then we will use the
following short hand notations.

\begin{itemize}
\item $\operatorname*{dp}(P)=\operatorname*{dp}(\operatorname*{cm}(P))$,

\item ${\operatorname*{pcdp}}(P)={\operatorname*{pcdp}}(\operatorname*{cm}%
(P)).$
\end{itemize}
\end{notation}

\begin{example}
Let $P=(P_{1},P_{2},P_{3})$ be as in Example \ref{example:P-cm(P)}. Thus

\begin{itemize}
\item ${\operatorname*{dp}}(P) ={\operatorname*{dp}}({\operatorname*{cm}}(P))
={\operatorname*{dp}}\left[
\begin{array}
[c]{cccc}%
b_{03} & b_{02} & b_{01} & b_{00}\\
b_{13} & b_{12} & b_{11} & b_{10}\\
& b_{22} & b_{21} & b_{20}%
\end{array}
\right]  = c_{1}x+c_{0} $, where
\begin{align*}
c_{1}=\det\left[
\begin{array}
[c]{cc|c}%
b_{03} & b_{02} & b_{01}\\
b_{13} & b_{12} & b_{11}\\
& b_{22} & b_{21}%
\end{array}
\right]  , \ \ c_{0}={\det}\left[
\begin{array}
[c]{cc|c}%
b_{03} & b_{02} & b_{00}\\
b_{13} & b_{12} & b_{10}\\
& b_{22} & b_{20}%
\end{array}
\right]
\end{align*}

\item ${\operatorname*{pcdp}}(P)={\operatorname*{pcdp}}(\operatorname*{cm}%
(P))=c_{1}=\det\left[
\begin{array}
[c]{cc|c}%
b_{03} & b_{02} & b_{01}\\
b_{13} & b_{12} & b_{11}\\
& b_{22} & b_{21}%
\end{array}
\right]  $
\end{itemize}
\end{example}

\medskip

\noindent Next we recall the concept of subresultant for two univariate polynomials.

\begin{definition}
\label{def:sres_2polys} Let $F_{0},F_{1}\in\mathcal{Z}[x]$ with $\deg
(F_{i})=d_{i}$ and $d_{0}\leq d_{1}$. Let $0<k\leq d_{0}$.

\begin{itemize}
\item The $k$-\emph{subresultant} of $F_{0}$ and $F_{1}$, written as
$R_{k}(F_{0},F_{1})$, is defined by
\[
R_{k}(F_{0},F_{1})={\operatorname*{dp}}(x^{d_{1}-(d_{0}-k)-1}F_{0}
,\ldots,x^{0}F_{0},x^{k-1}F_{1},\ldots,x^{0}F_{1})
\]

\item The \emph{principal} \emph{coefficient} of $R_{k}(F_{0},F_{1})$, written
as $r_{k}(F_{0},F_{1})$, is defined by
\[
r_{k}(F_{0},F_{1})=\operatorname*{coeff}\nolimits_{x^{d_{0}-k}}\left(
R_{k}\left(  F_{0},F_{1}\right)  \right)
\]

\end{itemize}

\noindent One can extend the above definition to the case when $k=0$. In this
case, it is required that $d_{0}\neq d_{1}$. Then
\[
R_{0}(F_{0},F_{1})=a_{0d_{0}}^{d_{1}-d_{0}-1}F_{0}
\]
where $a_{0d_{0}}$ is the leading coefficient of $F_{0}$.
\end{definition}

\begin{example}
Let
\begin{align*}
F_{0}  &  =a_{03}x^{3}+a_{02}x^{2}+a_{01}x+a_{00}\\
F_{1}  &  =a_{14}x^{4}+a_{13}x^{3}+a_{12}x^{2}+a_{11}x+a_{10}%
\end{align*}
Let $k=2$. Then%
\[
P=\left(  x^{2}F_{0},x^{1}F_{0},x^{0}F_{0},x^{1}F_{1},x^{0}F_{1}\right)
\]
Thus

\begin{itemize}
\item The $2$-\emph{subresultant} of $F_{0}$ and $F_{1},$ written as
$R_{2}(F_{0},F_{1})$, is
\begin{align*}
R_{2}(F_{0},F_{1})  &  ={\operatorname*{dp}}(P)\\
&  ={\operatorname*{dp}}({\operatorname*{cm}}(P))\\
&  ={\operatorname*{dp}}\left[
\begin{array}
[c]{cccccc}%
a_{03} & a_{02} & a_{01} & a_{00} &  & \\
& a_{03} & a_{02} & a_{01} & a_{00} & \\
&  & a_{03} & a_{02} & a_{01} & a_{00}\\\hline
a_{14} & a_{13} & a_{12} & a_{11} & a_{10} & \\
& a_{14} & a_{13} & a_{12} & a_{11} & a_{10}%
\end{array}
\right] \\
&  =\det\left[
\begin{array}
[c]{ccccc}%
a_{03} & a_{02} & a_{01} & a_{00} & \\
& a_{03} & a_{02} & a_{01} & a_{00}\\
&  & a_{03} & a_{02} & a_{01}\\\hline
a_{14} & a_{13} & a_{12} & a_{11} & a_{10}\\
& a_{14} & a_{13} & a_{12} & a_{11}%
\end{array}
\right]  x+{\det}\left[
\begin{array}
[c]{ccccc}%
a_{03} & a_{02} & a_{01} & a_{00} & \\
& a_{03} & a_{02} & a_{01} & \\
&  & a_{03} & a_{02} & a_{00}\\\hline
a_{14} & a_{13} & a_{12} & a_{11} & \\
& a_{14} & a_{13} & a_{12} & a_{10}%
\end{array}
\right]
\end{align*}

\item The \emph{principal} \emph{coefficient} of $R_{2}(F_{0},F_{1})$, written
as $r_{2}(F_{0},F_{1})$, is
\begin{align*}
r_{2}(F_{0},F_{1})  &  =\operatorname*{coeff}\nolimits_{x^{1}}\left(
R_{2}\left(  F_{0},F_{1}\right)  \right) \\
&  =\det\left[
\begin{array}
[c]{ccccc}%
a_{03} & a_{02} & a_{01} & a_{00} & \\
& a_{03} & a_{02} & a_{01} & a_{00}\\
&  & a_{03} & a_{02} & a_{01}\\\hline
a_{14} & a_{13} & a_{12} & a_{11} & a_{10}\\
& a_{14} & a_{13} & a_{12} & a_{11}%
\end{array}
\right]
\end{align*}

\end{itemize}
\end{example}

\begin{remark}
\

\begin{itemize}
\item Note that we are using a new indexing for subresultants. For instance,
$R_{k}(F_{0},F_{1})$ in the new indexing would have been indexed as
$R_{d_{0}-k}(F_{0},F_{1})$ in the classical indexing.

\item Note that we are using the terminologies \textquotedblleft subresultant"
and \textquotedblleft principal coefficient" as convention. In some other
literatures, readers may also see the terminologies \textquotedblleft
subresultant" and \textquotedblleft principal subresultant coefficient" (e.g.,
\cite{Collins:1976a,1993_Mishra}), or \textquotedblleft polynomial
subresultant" and \textquotedblleft scalar subresultant" (e.g.,
\cite{2003_von_zur_Gathen_Lucking}), \textquotedblleft subresultant
polynomial" and \textquotedblleft subresultant" (e.g., \cite{DAndrea:2006})
as their alternatives.
\end{itemize}
\end{remark}

Habicht discovered two intrinsic relationships among subresultants of two
polynomials in \cite{Habicht:1948}. The first is the similarity of a
subresultant with the pseudo-remainder of its two consecutive subresultants
and the second is the similarity of a subresultant with the subresultant of
two others. It should be pointed out that the first result can be viewed as a
specialization of the second one. In \cite{Hong_Yang-2023}, the authors
presented an analogy of the first result for several polynomials (which is not
a generalization of the classical result). In this paper, we will generalize
the second result to several polynomials. Therefore, we reproduce the second
result below.

\begin{theorem}
[Habicht's Theorem \cite{Habicht:1948}]\label{thm:structure_2poly}
We have
\begin{equation}
\label{eq:Habicht_2poly}r_{w_{0}}^{\epsilon}(F_{0},F_{1})\;R_{u}(F_{0}%
,F_{1})=R_{v}(R_{w_{0}}(F_{0},F_{1}),R_{w_{1}}(F_{0},F_{1}))
\end{equation}
if $u\leq d_{0},v,w_{0},w_{1}, \epsilon$ satisfy the following conditions:%

\[
\left\{
\begin{array}
[c]{rcl}%
w_{1} & = & w_{0}+1\\
v & \geq & 0\\
u & = & w_{0}+v\\
\epsilon & = & 2v-2
\end{array}
\right.
\]

\end{theorem}


\subsection{Review on subresultants of several polynomials}

For generalizing the classical Habicht's theorem to more than two polynomials,
we need the notions/notations of generalized subresultants for several
polynomials. In \cite{Hong_Yang-2023}, the authors present a natural extension
of the classical subresultant for two polynomials to multiple polynomials. For
the readers' convenience, we reproduce them here.

\begin{notation}
\label{notation1} \

\begin{itemize}
\item $d=(d_{0},\ldots,d_{n})\in\mathbb{N}^{n+1}$;

\item $a_{i}=\left(  a_{i0},\ldots,a_{id_{i}}\right)  $ be indeterminates (parameters);

\item $F=(F_{0},\ldots,F_{n})$ where $F_{i}=\sum_{j=0}^{d_{i}}a_{ij}x^{j}%
\in\mathbb{Z}[a_{i}][x]$;

\item $P(d_{0},n)=\{(\delta_{1},\ldots,\delta_{n})\in\mathbb{N}^{n}%
:\,|\delta|=\delta_{1}+\cdots+\delta_{n}\leq d_{0}\}$;

\item $\mathcal{F}_{k}=x^{\delta_{k}-1}F_{k},\ldots,x^{0}F_{k}$ where
$\delta_{k}\in\mathbb{N}$;

\item $c\left(  \delta\right)  =\#\operatorname{col}\operatorname{cm}\left(
\mathcal{F}_{1},\ldots,\mathcal{F}_{n}\right)  . $
\end{itemize}
\end{notation}

\begin{example}
Let $d=(3,3,4)$. Then
\[
P(d_{0}%
,n)=\{\ (3,0),\ (2,1),\ (1,2),\ (0,3),\ (2,0),\ (1,1),\ (0,2),\ (1,0),\ (0,1),\ (0,0)\ \}
\]
Choose $\delta=(1,1)\in P(3,2)$. We have
\[
\mathcal{F}_{1}=x^{0}F_{1},\quad\mathcal{F}_{2}=x^{0}F_{2}
\]
Thus
\begin{align*}
\operatorname{cm}\left(  \mathcal{F}_{1},\mathcal{F}_{2}\right)   &
=\operatorname{cm}\ (x^{0}F_{1},x^{0}F_{2}) =%
\begin{bmatrix}
& a_{13} & a_{12} & a_{11} & a_{10}\\
a_{24} & a_{23} & a_{22} & a_{21} & a_{20}%
\end{bmatrix}
\\
c\left(  \delta\right)   &  =\#\operatorname{col}%
\begin{bmatrix}
& a_{13} & a_{12} & a_{11} & a_{10}\\
a_{24} & a_{23} & a_{22} & a_{21} & a_{20}%
\end{bmatrix}
=5
\end{align*}

\end{example}

\begin{definition}
[Subresultant]\label{def:sres_npoly} Let $\delta\in P(d_{0},n)$.

\begin{itemize}
\item The $\delta$-\emph{subresultant} of $F$, written as $R_{\delta}\left(
F\right)  $, is defined by
\[
R_{\delta}\left(  F\right)  ={\operatorname*{dp}}\operatorname{cm}\left(
\mathcal{F}_{0},\ldots,\mathcal{F}_{n}\right)
\]
where again%
\begin{equation}
\label{eqs:delta0}\delta_{0}=\left\{
\begin{array}
[c]{ll}%
c\left(  \delta\right)  -d_{0} & \text{if\ \ }c\left(  \delta\right)  \geq
d_{0}\\
1 & \text{else}%
\end{array}
\right.
\end{equation}

\item The \emph{principal} \emph{coefficient} of $R_{\delta}\left(  F\right)
$, written as $r_{\delta}(F)$, is defined by
\[
r_{\delta}(F)=\operatorname*{coeff}\nolimits_{x^{d_{0}-\left\vert
\delta\right\vert }}\left(  R_{\delta}\left(  F\right)  \right)
\]

\end{itemize}
\end{definition}

\begin{remark}
In the above, the particular expression for $\delta_{0}$ is chosen because it
naturally extends the formulation of subresultants for two polynomials.
Roughly speaking, such choice of $\delta_{0}$ makes the submatrix of
$\operatorname{cm}\left(  \mathcal{F}_{0},\ldots,\mathcal{F}_{n}\right)  $
involving the coefficients of $F_{0}$ the widest block while keeping the size
of the matrix $\operatorname{cm}\left(  \mathcal{F}_{0},\ldots,\mathcal{F}%
_{n}\right)  $ as small as possible.
\end{remark}

\begin{example}
Let $d=(3,3,4)$ and $\delta=\left(  1,1\right)  $. Note%
\[
c\left(  \delta\right)  =\#\operatorname{col}\operatorname{cm}\left(
\mathcal{F}_{1},\mathcal{F}_{2}\right)  =5\geq d_{0},\ \ \ \delta_{0}=5-3=2
\]
Therefore
\begin{align*}
R_{(1,1)}(F)  &  ={\operatorname*{dp}}%
\begin{bmatrix}
a_{03} & a_{02} & a_{01} & a_{00} & \\
& a_{03} & a_{02} & a_{01} & a_{00}\\
& a_{13} & a_{12} & a_{11} & a_{10}\\
a_{24} & a_{23} & a_{22} & a_{21} & a_{20}%
\end{bmatrix}
=\det%
\begin{bmatrix}
a_{03} & a_{02} & a_{01} & a_{00}\\
& a_{03} & a_{02} & a_{01}\\
& a_{13} & a_{12} & a_{11}\\
a_{24} & a_{23} & a_{22} & a_{21}%
\end{bmatrix}
x+\det%
\begin{bmatrix}
a_{03} & a_{02} & a_{01} & \\
& a_{03} & a_{02} & a_{00}\\
& a_{13} & a_{12} & a_{10}\\
a_{24} & a_{23} & a_{22} & a_{20}%
\end{bmatrix}
\\
r_{(1,1)}(F)  &  =\det%
\begin{bmatrix}
a_{03} & a_{02} & a_{01} & a_{00}\\
& a_{03} & a_{02} & a_{01}\\
& a_{13} & a_{12} & a_{11}\\
a_{24} & a_{23} & a_{22} & a_{21}%
\end{bmatrix}
\end{align*}

\end{example}

\section{Main Result}

\label{sec:main_result} In this section, we describe a generalization of
Habicht's theorem for several polynomials. For this purpose, we need the
following notation.

\begin{notation}
Let $e_{i}\in\{0,1\}^{n}$ be the $i$-th unit vector of length $n$, that is,
the vector whose $i$-th element is 1 and the remaining elements are all zeros.
We choose the convention that $e_{0}= \left(  0,\ldots,0\right)  $.
\end{notation}

\begin{assumption}
From now on, we will assume that $d_{0}\leq d_{1},\ldots,d_{n}$ and $F_{0}$ is
monic, i.e., $a_{0d_{0}}=1$.
\end{assumption}

\begin{theorem}
[Main Result]\label{thm:main} We have
\begin{equation}
r_{w_{0}}^{\epsilon}(F_{0},\ldots,F_{n})\ \;R_{u}(F_{0},\ldots,F_{n}%
)\ \ \ =\ \ \ R_{v}(R_{w_{0}}(F_{0},\ldots,F_{n}),\ldots,R_{w_{n}}%
(F_{0},\ldots,F_{n})) \label{eq:main_result}%
\end{equation}
if $u\in P\left(  d_{0},n\right)  ,v,w_{0},\ldots,w_{n},\epsilon$ satisfy the
following conditions
\[
\left\{
\begin{array}
[c]{rcl}%
w_{1} & = & w_{0}+e_{1}\\
& \vdots & \\
w_{n} & = & w_{0}+e_{n}\\
v & = & (k,\ldots,k)+e_{i}\\
u & = & w_{0}+v\\
\epsilon & = & |v+e_{i}|+k-2
\end{array}
\right.
\]
for some $k\geq1$ and $0\leq i\leq n$.
\end{theorem}

\begin{example}
Let $d=(5,5,6)$.

\begin{itemize}
\item Let $w_{0}=(1,1)$, $k=1$ and $i=0$. Then
\[
\omega_{1}=(2,1),\ \ \omega_{2}=(1,2),\ \ v=(1,1),\ \ u=(2,2),\ \ \epsilon
=|(1,1)|+1-2=1
\]
Thus we have
\[
r_{(1,1)}^{1}(F)\ R_{(2,2)}(F)\ \ \ =\ \ R_{(1,1)} \ (\ R_{(1,1)}%
(F),\;R_{(2,1)}(F),\;R_{(1,2)}(F)\ )
\]

\item Let $w_{0}=(1,1)$, $k=1$ and $i=1$. Then
\[
\omega_{1}=(2,1),\ \ \omega_{2}=(1,2),\ \ v=(2,1),\ \ u=(3,2),\ \ \epsilon
=|(3,1)|+1-2=3
\]
Thus we have
\[
r_{(1,1)}^{3}(F)\ R_{(3,2)}(F)\ \ \ =\ \ R_{(2,1)} \ (\ R_{(1,1)}%
(F),\;R_{(2,1)}(F),\;R_{(1,2)}(F)\ )
\]

\item Let $w_{0}=(1,1)$, $k=1$ and $i=2$. Then
\[
\omega_{1}=(2,1),\ \ \omega_{2}=(1,2),\ \ v=(1,2),\ \ u=(2,3),\ \ \epsilon
=|(1,3)|+1-2=3
\]
Thus we have
\[
r_{(1,1)}^{3}(F)\ R_{(2,3)}(F)\ \ \ =\ \ R_{(1,2)} \ (\ R_{(1,1)}%
(F),\;R_{(2,1)}(F),;R_{(1,2)}(F)\ )
\]

\end{itemize}
\end{example}


\section{Relation to the Classical Habicht's Theorem}

\label{sec:remarks} After setting $n=1$ in Theorem \ref{thm:main}, the theorem
reduces to the following:
\[
r_{w_{0}}^{\epsilon}(F_{0},F_{1})\;\ R_{u}(F_{0},F_{1})\ \ =\ \ R_{v}%
(R_{w_{0}}(F_{0},F_{1}),R_{w_{1}}(F_{0},F_{1}))
\]
if $u\in P\left(  d_{0},1\right)  (\text{i.e.,\ }u\leq d_{0}),v,w_{0}%
,w_{1},\epsilon$ satisfy the following conditions {
\begin{equation}
\left\{
\begin{array}
[c]{rcl}%
w_{1} & = & w_{0}+1\\
v & = & k+e_{i}\\
u & = & w_{0}+v\\
\epsilon & = & |v+e_{i}|+k-2
\end{array}
\right.  \label{eqs:condition_n=2}%
\end{equation}
for some $k\geq1$ and $0\leq i\leq1$. Let $i=0$.} Then the the condition
\eqref{eqs:condition_n=2} simplifies to
\[
\left\{
\begin{array}
[c]{rcl}%
w_{1} & = & w_{0}+1\\
v & = & k\\
u & = & w_{0}+k\\
\epsilon & = & 2k-2
\end{array}
\right.
\]
which is equivalent to
\[
\left\{
\begin{array}
[c]{rcl}%
w_{1} & = & w_{0}+1\\
v\ \  & \geq & \ \ 1\\
u & = & w_{0}+v\\
\epsilon & = & 2v-2
\end{array}
\right.
\]
as in Theorem \ref{thm:structure_2poly}. Thus we conclude that the classical
Habicht's theorem is indeed a specialization of the generalized version
presented in this paper.

In the remaining part of this section, we make a further analysis on the
similarity between the generalized Habicht's theorem (Theorem \ref{thm:main})
and the classical one (Theorem \ref{thm:structure_2poly}), which gives readers
a deeper understanding on the content of the generalized Habicht's theorem.
For the sake of simplicity, when $F$ is clear from the context, we can
abbreviate $R_{\delta}\left(  F\right)  $ and $r_{\delta}\left(  F\right)  $
as $R_{\delta}$ and $r_{\delta}$, respectively.

First, we note that the relationship \eqref{eq:Habicht_2poly} in the classical
Habicht's theorem (Theorem \ref{thm:structure_2poly}) can be illustrated as
\begin{align*}
\left[
\begin{array}
[c]{c}%
R_{w_{0}+e_{0}}\\
R_{w_{0}+e_{1}}%
\end{array}
\right]  \ \ \Longrightarrow\ \ \left[
\begin{array}
[c]{c}%
R_{w_{0}+1+e_{1}}%
\end{array}
\right]  \ \ \cdots\ \ \left[
\begin{array}
[c]{c}%
R_{w_{0}+k+e_{1}}%
\end{array}
\right]
\end{align*}
\noindent This diagram is interpreted as: $R_{w_{0}+j+e_{1}}$ for
$j=1,\ldots,k$ can be computed from $R_{w_{0}+e_{0}}$ and $R_{w_{0}+e_{1}}$.
Note that $R_{w_{0}+(j-1)+e_{1}}=R_{w_{0}+j+e_{0}}$ for $j=1,\ldots,k$. Thus,
the above relationship can be equivalently converted to the following:%

\begin{align*}
\left[
\begin{array}
[c]{c}%
R_{w_{0}+e_{0}}\\
R_{w_{0}+e_{1}}%
\end{array}
\right]  \ \ \Longrightarrow\ \ \left[
\begin{array}
[c]{c}%
R_{w_{0}+1+e_{0}}\\
R_{w_{0}+1+e_{1}}%
\end{array}
\right]  \ \ \cdots\ \ \left[
\begin{array}
[c]{c}%
R_{w_{0}+k+e_{0}}\\
R_{w_{0}+k+e_{1}}%
\end{array}
\right]
\end{align*}

\noindent This diagram is interpreted as: $R_{w_{0}+j+e_{i}}$ for $i=0,1$ and
$j=1,\ldots,k$ can be computed from $R_{w_{0}+e_{0}}$ and $R_{w_{0}+e_{1}}$.

The same pattern can be found in the generalized Habicht's theorem. More
explicitly, let $d=(d_{0},\ldots,d_{n})$ where $d_{0}=\min_{0\le i\le n}d_{i}$
and $k^{\prime}=(k,\ldots,k)$ for $k\ge1$, then $v\in\{k^{\prime}+e_{i}:
i=0,\ldots,n\}$. The relationship \eqref{eq:main_result} in the generalized
Habicht's theorem (Theorem \ref{thm:main}) can be illustrated as%

\begin{align*}
\left[
\begin{array}
[c]{c}%
R_{w_{0}+e_{0}}\\
R_{w_{0}+e_{1}}\\
\vdots\\
R_{w_{0}+e_{n}}%
\end{array}
\right]  \ \ \Longrightarrow\ \ \left[
\begin{array}
[c]{c}%
R_{w_{0}+1^{\prime}+e_{0}}\\
R_{w_{0}+1^{\prime}+e_{1}}\\
\vdots\\
R_{w_{0}+1^{\prime}+e_{n}}%
\end{array}
\right]  \ \ \cdots\ \ \left[
\begin{array}
[c]{c}%
R_{w_{0}+k^{\prime}+e_{0}}\\
R_{w_{0}+k^{\prime}+e_{1}}\\
\vdots\\
R_{w_{0}+k^{\prime}+e_{n}}%
\end{array}
\right]
\end{align*}

\noindent The diagram is again interpreted as: $R_{w_{0}+j^{\prime}+e_{i}}$
for $i=0,\ldots,n$ and $j=1,\ldots,k$ can be computed from $R_{w_{0}+e_{0}%
},\ldots,R_{w_{0}+e_{n}}$.

To have a better understanding of the similarity between the classical
Habicht's theorem and the generalized version, we introduce the concepts of
cluster and cluster chain of adjacent subresultants, which might be helpful
for identifying more relationships among subresultants.

\begin{definition}
We call $(R_{w_{0}},R_{w_{0}+e_{1}},\ldots,R_{w_{0}+e_{n}})$ a \emph{cluster
of subresultants} for $F$, denoted by $C_{w_{0}}$. Then $\ldots,C_{w_{0}%
},C_{w_{0}+1^{\prime}},\ldots$ is called the \emph{cluster chain of
subresultants}.
\end{definition}

With the above definition, we re-interpret the generalized Habicht's theorem
as follows. Given a cluster~$C_{w_{0}}$ of subresultants, we can compute all
the subresultants in the clusters in the right of $C_{w_{0}}$ along the
cluster chain. Note that the degrees of subresultants are decreasing along to
the right. With the degree becoming smaller, the subresultants tend to become
complicated because the determinant polynomials they are obtained from are
with higher order. Thus in the classical subresultants, one often computes
subresultants with lower degree (corresponding to higher-order determinant
polynomials) from those with higher degree (corresponding to lower-order
determinant polynomials).

We hope that this interpretation would have an impact on establishing inherent
relationships between subresultants of multiple polynomials and
pseudo-remainders, which would, in turn, lead to the efficient computation of
subresultants. For instance, in what follows, we apply the generalized
Habicht's theorem repeatedly to find further relations among subresultants.

\begin{example}
Let $d=\left(  5,5,6\right)  $. We would like to reduce $R_{(3,2)}(F)$ to
$R_{(\ast,0)}(F)$ or $R_{(0,\ast)}(F)$ by repeatedly applying Theorem
\ref{thm:main}. There are several approaches to achieve this. We will
illustrate two approaches.

\begin{enumerate}
\item Approach A.

\begin{enumerate}
\item $r_{(2,1)}\left(  F\right)  \ R_{(3,2)}(F)=R_{(1,1)}(R_{(2,1)}\left(
F\right)  ,R_{(3,1)}\left(  F\right)  ,R_{(2,2)}\left(  F\right)  )$ using
$w_{0}=(2,1)$, $k=1$ and $i=0$.

\item $r_{(1,0)}\left(  F\right)  \ R_{(2,1)}(F)=R_{(1,1)}(R_{(1,0)}\left(
F\right)  ,R_{(2,0)}\left(  F\right)  ,R_{(1,1)}\left(  F\right)  )$ using
$w_{0}=(1,0)$, $k=1$ and $i=0$.

$r_{(2,0)}\left(  F\right)  \ R_{(3,1)}\left(  F\right)  =R_{(1,1)}%
(R_{(2,0)}\left(  F\right)  ,R_{(3,0)}\left(  F\right)  ,R_{(2,1)}\left(
F\right)  )$ using $w_{0}=(2,0)$, $k=1$ and $i=0$.

$r_{(1,1)}\left(  F\right)  \ R_{(2,2)}\left(  F\right)  =R_{(1,1)}%
(R_{(1,1)}\left(  F\right)  ,R_{(2,1)}\left(  F\right)  ,R_{(1,2)}\left(
F\right)  )$ using $w_{0}=(1,1)$, $k=1$ and $i=0$.

\item $r_{(0,0)}\left(  F\right)  \ R_{(1,1)}\left(  F\right)  =R_{(1,1)}%
(R_{(0,0)}\left(  F\right)  ,R_{(1,0)}\left(  F\right)  ,R_{(0,1)}\left(
F\right)  )$ using $w_{0}=(0,0)$, $k=1$ and $i=0$.

$r_{(0,1)}\left(  F\right)  \ R_{(1,2)}\left(  F\right)  =R_{(1,1)}%
(R_{(0,1)}\left(  F\right)  ,R_{(1,1)}\left(  F\right)  ,R_{(0,2)}\left(
F\right)  )$ using $w_{0}=(0,1)$, $k=1$ and $i=0$.
\end{enumerate}

Thus we have reduced $R_{(3,2)}(F)$ to
\[
R_{(0,0)}(F),\ R_{(0,1)}(F),\ R_{(0,2)}(F),\ R_{(1,0)}(F),\ R_{(2,0)}%
(F),\ R_{(3,0)}(F)
\]
repeatedly using $R_{(1,1)}.$ Schematically, we have
\[%
\begin{array}
[c]{ccccccc}%
02 & \longrightarrow^{1} & 12 & \longrightarrow^{1} & 22 & \longrightarrow^{1}
& 32\\
& \nearrow^{1} & \uparrow & \nearrow^{1} & \uparrow^{1} & \nearrow^{1} &
\uparrow^{1}\\
01 & \longrightarrow^{1} & 11 & \longrightarrow^{1} & 21 & \longrightarrow^{1}
& 31\\
& \nearrow^{1} & \uparrow^{1} & \nearrow^{1} & \uparrow^{1} & \nearrow^{1} &
\uparrow^{1}\\
00 &  & 10 &  & 20 &  & 30
\end{array}
\]

\item Approach B.

\begin{enumerate}
\item $r_{(1,0)}^{4}\left(  F\right)  \ R_{(3,2)}(F)=R_{(2,2)}(R_{(1,0)}%
\left(  F\right)  ,R_{(2,0)}\left(  F\right)  ,R_{(1,1)}\left(  F\right)  )$
using $w_{0}=(1,0)$, $k=2$ and $i=0$.

\item $r_{(0,0)}\left(  F\right)  \ R_{(1,1)}\left(  F\right)  =R_{(1,1)}%
(R_{(0,0)}\left(  F\right)  ,R_{(1,0)}\left(  F\right)  ,R_{(0,1)}\left(
F\right)  )$ using $w_{0}=(0,0)$, $k=1$ and $i=0$.
\end{enumerate}

Thus we have reduced $R_{(3,2)}(F)$ to
\[
R_{(0,0)}(F),\ R_{(0,1)}(F),\ R_{(1,0)}(F),R_{(2,0)}(F)
\]
using $R_{(2,2)}\ $and $R_{(1,1)}.$ Schematically, we have
\[%
\begin{array}
[c]{ccccc}%
01 & \longrightarrow^{1} & 11 & \longrightarrow^{2} & 32\\
& \nearrow^{1} & \uparrow^{1} & \nearrow^{2} & \uparrow^{2}\\
00 &  & 10 &  & 20
\end{array}
\]

\end{enumerate}

\noindent The above two approaches provide two ways for reducing subresultants
of high order to nested subresultants of low orders. Their results differ in
two aspects.

\begin{itemize}
\item The subresultants in the inner layers involved in Approach B are of
lower order compared with those involved in Approach A;

\item The subresultants in the outer layers involved in Approach B are of
higher order compared with those involved in Approach A.
\end{itemize}
\end{example}

\section{Proof of the Generalized Habicht's Theorem (Theorem \ref{thm:main})}

\label{sec:proof} This section is devoted to proving the generalized Habicht's
theorem. The proof will be given by induction on $k$, starting from $k=1$ as
the induction base. The proof for the induction base is given in
Subsection~\ref{subsec:base} and the proof for the induction step is given in
Subsection~\ref{subsec:inductive}.
Those proofs depend on certain  properties of determinant polynomials. Thus we  begin by introducing and proving those properties  in Subsection~\ref{ssec:detp_properties}  before we carry out induction.

\subsection{Some useful properties of determinant polynomials}
\label{ssec:detp_properties} In this subsection, we derive or recall several
properties of determinant polynomials, which will be used in the following two subsections.

\begin{proposition}
\label{prop:ci} Let $P=(P_{1},\ldots,P_{t})$ where $P_{i}=\sum_{0\le j\le
p_{i}}b_{ij}x^{j}$ and $p_{i}=\deg P_{i}$. Let $m=\max_{1\le i\le t} p_{i}$.
If $t \leq m+1$, then we have
\[
{\operatorname*{dp}}(P_{1},\ldots,P_{t})=\sum_{i=1}^{t}c_{i}P_{i}
\]
where
\begin{align*}
c_{i}=\left\{
\begin{array}
[c]{ll}%
1 & \text{if}\ \ t=1;\\
0 & \text{if}\ \ t>1 \ \wedge\ \underset{j\neq i}{\forall}p_{j}< p_{i};\\
(-1)^{t+i}{\operatorname*{pcdp}}(P_{1},\ldots,P_{i-1},P_{i+1},\ldots,P_{t}) &
\text{else}%
\end{array}
\right.
\end{align*}

\end{proposition}

\begin{proof}
If $t=1$, it is clear that ${\operatorname*{dp}}(P_{1})=c_{1}P_{1}$ where
$c_{1}=1$. Next we consider the case where $t>1$. By the multi-linearity of
determinant, we have
\begin{align*}
\operatorname*{dp}(P_{1},\dots,P_{t}) =\sum_{i=0}^{m-t+1}\det\left[
\setlength\arraycolsep{3pt}
\begin{array}
[c]{cccc}%
b_{1m} & \cdots & b_{1(m-t+2)} & b_{1i}\\
\vdots &  & \vdots & \vdots\\
b_{tm} & \cdots & b_{t(m-t+2)} & b_{ti}%
\end{array}
\right]  x^{i} =\det\left[
\begin{array}
[c]{cccc}%
b_{1m} & \cdots & b_{1(m-t+2)} & \sum\limits_{i=0}^{m-t+1}b_{1i}x^{i}\\
\vdots &  & \vdots & \vdots\\
b_{tm} & \cdots & b_{t(m-t+2)} & \sum\limits_{i=0}^{m-t+1}b_{ti}x^{i}%
\end{array}
\right]
\end{align*}
where $b_{ij}:=0$ for $j>p_{i}$. Adding the $i$-th column multiplied by
$x^{m+1-i}$ to the last column, we get
\begin{align*}
\operatorname*{dp}(P_{1},\dots,P_{t}) =\det\left[
\begin{array}
[c]{cccc}%
b_{1m} & \cdots & b_{1(m-t+2)} & \sum\limits_{i=0}^{m}b_{1i}x^{i}\\
\vdots &  & \vdots & \vdots\\
b_{tm} & \cdots & b_{t(m-t+2)} & \sum\limits_{i=0}^{m}b_{ti}x^{i}%
\end{array}
\right]  =\det\left[
\begin{array}
[c]{cccc}%
b_{1m} & \cdots & b_{1(m-t+2)} & P_{1}\\
\vdots &  & \vdots & \vdots\\
b_{tm} & \cdots & b_{t(m-t+2)} & P_{t}%
\end{array}
\right]
\end{align*}
Then the expansion of the matrix along the last column results in
\begin{align*}
{\operatorname*{dp}}(P_{1},\ldots,P_{t})=\sum_{i=1}^{t}(-1)^{t+i}\det
(M^{(i)})\cdot P_{i}%
\end{align*}
where
\begin{align}
\label{eq:M_i}M^{(i)}=\left[
\begin{array}
[c]{ccc}%
b_{1m} & \cdots & b_{1(m-t+2)}\\
\vdots &  & \vdots\\
b_{(i-1)m} & \cdots & b_{(i-1)(m-t+2)}\\
b_{(i+1)m} & \cdots & b_{(i+1)(m-t+2)}\\
\vdots &  & \vdots\\
b_{tm} & \cdots & b_{t(m-t+2)}%
\end{array}
\right]
\end{align}

Let $c_{i}=(-1)^{t+i}\det(M^{(i)})$, then ${\operatorname*{dp}}(P_{1}%
,\ldots,P_{t})=\sum_{i=1}^{t}c_{i}P_{i}$. Now we consider the following two
cases for $c_{i}$, depending on whether $\underset{j\neq i}{\forall}p_{j}<
p_{i}$ holds or not.

\begin{itemize}
\item[C1:] $\underset{j\neq i}{\forall}p_{j}< p_{i}$.

In this case, we have $b_{jm}=0$ for $j=1,\ldots,i-1,i+1,\ldots,t$. By
\eqref{eq:M_i},
\begin{align*}
c_{i}=\sum_{i=1}^{t}(-1)^{t+i}\det(M^{(i)})=0
\end{align*}

\item[C2:] $\underset{j\neq i}{\exists}p_{j}\geq p_{i}$. In this case,
$\det(M^{(i)})={\operatorname*{pcdp}}\left(  P_{1},\ldots,P_{i-1}%
,P_{i+1},\ldots,P_{t}\right)  $. Then
\begin{align*}
c_{i}=\sum_{i=1}^{t}(-1)^{t+i}\det(M^{(i)})=(-1)^{t+i}{\operatorname*{pcdp}%
}\left(  P_{1},\ldots,P_{i-1},P_{i+1},\ldots,P_{t}\right)
\end{align*}

\end{itemize}

The proof is completed.
\end{proof}

\medskip\noindent The following result was first presented in
\cite{Hong_Yang-2023}. It is a specialization of Proposition \ref{prop:ci}
when $P$ is specialized with $\left(  x^{\delta_{0}-1}F_{0},\ldots,x^{0}%
F_{0},\ldots,x^{\delta_{n}-1}F_{n},\ldots,x^{0}F_{n}\right)  $.

\begin{corollary}
\label{cor:dp} Let $\delta=(\delta_{1},\ldots,\delta_{n})\in P(n,d_{0})$ and
$F=(F_{0},\ldots,F_{n})$ in Notation \ref{notation1}. Then we have
\[
R_{\delta}(F)=\sum_{i=0}^{n}\sum_{j=0}^{\delta_{i}-1}c_{ij}x^{j}F_{i}
\]
where $c_{ij}\in\mathcal{Z}[a_{0},\ldots,a_{n}]$. In particular, when $i>0$
and $\delta_{i}\ne0$,
\begin{align*}
c_{i,\delta_{i}-1}=(-1)^{\sigma_{i}+1}r_{\delta-e_{i}}(F)
\end{align*}
where $\sigma_{i}=\delta_{i}+\cdots+\delta_{n}$.


\end{corollary}


\begin{lemma}
\label{lemma:002} Let $P=(P_{1},\ldots,P_{t_{1}})$, $Q=(Q_{1},\ldots,Q_{t_{2}%
})$ with $\deg P_{i}=p_{i}$ and $\deg Q_{i}=q_{i}$ be such that $t_{1}%
,t_{2}\ge1$ and $m_{1}=\max\limits_{1\le i\le t_{1}} p_{i}\ge m_{2}%
=\max\limits_{1\le i\le t_{2}} q_{i}$. If $t_{1}+t_{2}\leq m_{1}+1$ and
$m_{2}\le m_{1}-t_{1}+1$, then we have the following:%

\[
\operatorname*{dp}(P,Q)=\left\{
\begin{array}
[c]{ll}%
\text{$(1)$}\ \ \operatorname*{dp}(\operatorname*{dp}(P),Q) & \text{if}%
\ \ m_{2}=m_{1}-t_{1}+1;\\
\text{$(2)$}\ \ {\operatorname*{pcdp}}(P)\cdot\operatorname*{dp}(Q) &
\text{if}\ \ m_{2}=m_{1}-t_{1}\ \ \text{or}\ \ m_{2}< m_{1}-t_{1} \wedge
t_{2}= 1;\\
\text{$(3)$}\ \ 0 & \text{else}%
\end{array}
\right.
\]

\end{lemma}

\begin{proof}
Assume $P_{i}=\sum_{0\le j\le p_{i}}b_{ij}x^{j}$, $Q_{i}=\sum_{0\le j\le
q_{i}}b_{ij}^{*}x^{j}$. Let $M^{(1)}={\operatorname*{cm}}(P_{1},\ldots
,P_{t_{1}},Q_{1},\ldots,Q_{t_{2}})$, that is
\begin{align*}
M^{(1)}=\left[
\begin{array}
[c]{cccccc}%
b_{1m_{1}} & \cdots & b_{1(m_{1}-t_{1}+1)} & b_{1(m_{1}-t_{1})} & \cdots &
b_{10}\\
\vdots &  & \vdots & \vdots &  & \vdots\\
b_{t_{1}m_{1}} & \cdots & b_{t_{1}(m_{1}-t_{1}+1)} & b_{t_{1}(m_{1}-t_{1})} &
\cdots & b_{t_{1}0}\\
&  & b_{1(m_{1}-t_{1}+1)}^{*} & b_{1(m_{1}-t_{1})}^{*} & \cdots & b_{10}^{*}\\
&  & \vdots & \vdots &  & \vdots\\
&  & b_{t_{2}(m_{1}-t_{1}+1)}^{*} & b_{t_{2}(m_{1}-t_{1})}^{*} & \cdots &
b_{t_{2}0}^{*}%
\end{array}
\right]
\end{align*}
where $b_{ij}:=0$ when $j>p_{i}$ and $b_{ij}^{*}:=0$ when $j>q_{i}$. Next we
simplify $\operatorname*{dp} (M^{(1)})$. For this purpose, let $U_{1}%
=\operatorname*{dp}(P_{1},\ldots,P_{t_{1}})$. By Proposition \ref{prop:ci}, we
have
\begin{align*}
U_{1}=\operatorname*{dp}(P_{1},\ldots,P_{t_{1}})=c_{1}P_{1}+\cdots+c_{t_{1}%
}P_{t_{1}}%
\end{align*}
where
\begin{align*}
c_{i}=\left\{
\begin{array}
[c]{ll}%
1 & \text{if}\ \ t_{1}=1;\\
0 & \text{if}\ \ t_{1}>1\ \wedge\underset{j\neq i}{\forall}p_{j}< p_{i};\\
(-1)^{t_{1}+i}{\operatorname*{pcdp}}(P_{1},\ldots,P_{i-1},P_{i+1}%
,\ldots,P_{t_{1}}) & \text{else}%
\end{array}
\right.
\end{align*}
\noindent Now we proceed to analyze $\operatorname*{dp}(M^{(1)})$ separately
under the two conditions of $t_{1}>1$ and $t_{1}=1$.

\begin{itemize}
[leftmargin=15mm]

\item[\textbf{Case(i)}] : $t_{1}=1$.

In this case, $U_{1}=\operatorname*{dp}(P_{1})=P_{1}$. Then we have
\begin{align*}
\operatorname*{dp}(M^{(1)})=\operatorname*{dp}(P_{1},Q_{1},\ldots,Q_{t_{2}%
})=\operatorname*{dp}(U_{1},Q_{1},\ldots,Q_{t_{2}})
\end{align*}

\item[\textbf{Case(ii)}] $:t_{1}>1$.

Without loss of generality, we suppose that $p_{t_{1}}=\min\limits_{1\le i\le
t_{1}} p_{i}$, i.e., $c_{t_{1}}\neq0$. It follows that
\begin{align}
\label{equ:001}c_{t_{1}}\operatorname*{dp}(M^{(1)})  &  =\operatorname*{dp}%
(P_{1},\ldots,P_{t_{1}-1},c_{t_{1}}P_{t_{1}},Q_{1},\ldots,Q_{t_{2}%
})\nonumber\\
&  =\operatorname*{dp}(P_{1},\ldots,P_{t_{1}-1},U_{1},Q_{1},\ldots,Q_{t_{2}})
\end{align}
Consider $M^{(2)}=\operatorname*{cm}(P_{1},\ldots,P_{t_{1}-1},U_{1}%
,Q_{1},\ldots,Q_{t_{2}})$. We partition $M^{(2)}$ in the following way:
\begin{align*}
M^{(2)}=\left[
\begin{array}
[c]{cc}%
N^{(1)} & N^{(2)}\\
& N^{(3)}%
\end{array}
\right]
\end{align*}
where
\begin{align*}
N^{(1)}=\left[
\begin{array}
[c]{ccc}%
b_{1m_{1}} & \cdots & b_{1(m_{1}-t_{1}+2)}\\
\vdots &  & \vdots\\
b_{(t_{1}-1)m_{1}} & \cdots & b_{(t_{1}-1)(m_{1}-t_{1}+2)}\\
&  &
\end{array}
\right]  , N^{(2)}=\left[
\begin{array}
[c]{ccc}%
b_{1(m_{1}-t_{1}+1)} & \cdots & b_{10}\\
\vdots &  & \vdots\\
b_{(t_{1}-1)(m_{1}-t_{1}+1)} & \cdots & b_{(t_{1}-1)0}%
\end{array}
\right]
\end{align*}
and
\[
N^{(3)}=\operatorname*{dp}(U_{1},Q_{1},\ldots,Q_{t_{2}})=\left[
\begin{array}
[c]{ccc}%
u_{m_{1}-t_{1}+1} & \cdots & u_{0}\\
b_{1(m_{1}-t_{1}+1)}^{*} & \cdots & b_{10}^{*}\\
\vdots &  & \vdots\\
b_{t_{2}(m_{1}-t_{1}+1)}^{*} & \cdots & b_{t_{2}0}^{*}%
\end{array}
\right]
\]
where $u_{i}$ is the coefficient of $U_{1}$ in the term $x^{i}$. Next we show
that when $m_{2}\le m_{1}-t_{1}+1$, $\operatorname*{dp}(M^{(1)}%
)=\operatorname*{dp} (N^{(3)})$.

Let $M_{i}^{(2)}$, $N_{i}^{(1)}$, $N_{i}^{(2)}$ and $N_{i}^{(3)}$ be the
$i$-th column of $M^{(2)}$, $N^{(1)}$, $N^{(2)}$ and $N^{(3)}$, respectively.
By the definition of determinant polynomial, we have
\begin{align*}
\operatorname*{dp}(M^{(2)})  &  =\sum_{i=t_{1}+t_{2}}^{m_{1}+1}\det\left[
\begin{array}
[c]{cccc}%
M_{1}^{(2)} & \cdots & M_{t_{1}+t_{2}-1}^{(2)} & M_{i}^{(2)}%
\end{array}
\right]  \cdot x^{m_{1}+1-i}\\
&  =\sum_{i=t_{2}+1}^{m_{1}-t_{1}+2}\det\left[
\begin{array}
[c]{cccllll}%
N_{1}^{(1)} & \cdots & N_{t_{1}-1}^{(1)} & N_{1}^{(2)} & \cdots & N_{t_{2}%
}^{(2)} & N_{i}^{(2)}\\
&  &  & N_{1}^{(3)} & \cdots & N_{t_{2}}^{(3)} & N_{i}^{(3)}%
\end{array}
\right]  \cdot x^{m_{1}-t_{1}+2-i}%
\end{align*}
Since $N^{(1)}=%
\begin{bmatrix}
N_{1}^{(1)} & \cdots & N_{t_{1}-1}^{(1)}%
\end{bmatrix}
$ is a square matrix of order $t_{1}-1$, we have
\begin{align}
\label{equ:002}\operatorname*{dp}(M^{(2)})  &  =\det N^{(1)}\cdot\sum
_{i=t_{2}+1}^{m_{1}-t_{1}+2} \det\left[
\begin{array}
[c]{cccc}%
N_{1}^{(3)} & \cdots & N_{t_{2}}^{(3)} & N_{i}^{(3)}\\
&  &  &
\end{array}
\right]  x^{m_{1}-t_{1}+2-i}\nonumber\\
&  =\det(N^{(1)})\cdot\operatorname*{dp}(N^{(3)})
\end{align}
The substitution of \eqref{equ:002} into \eqref{equ:001} yields
\begin{align*}
c_{t_{1}}\cdot\operatorname*{dp}(M^{(1)})=\operatorname*{dp}(M^{(2)}%
)=\det(N^{(1)})\cdot\operatorname*{dp}(N^{(3)})
\end{align*}
Noting that $c_{t_{1}}={\operatorname*{pcdp}}(P_{1},\ldots,P_{t_{1}-1}%
)=\det(N^{(1)})$, we have
\begin{align*}
\operatorname*{dp}(M^{(1)})=\operatorname*{dp}(N^{(3)})
\end{align*}

\end{itemize}

\noindent Therefore, we can deduce that for both the scenarios where $t_{1}=1$
and $t_{1}>1$, the following equation holds:
\begin{align}
\label{lem:result1}\operatorname*{dp}(M^{(1)})=\operatorname*{dp}(U_{1}%
,Q_{1},\ldots,Q_{t_{2}})=\operatorname*{dp}(N^{(3)})
\end{align}

\medskip

\noindent Next, we consider the following four cases for $m_{2}\le m_{1}%
-t_{1}+1$ and specialize the above results for these cases.

\begin{itemize}
[leftmargin=15mm]

\item[Case (1):] $m_{2}=m_{1}-t_{1}+1$.

In this case, $N^{(3)}=\operatorname*{cm}(U_{1},Q_{1},\ldots,Q_{t_{2}%
})=\operatorname*{cm}(\operatorname*{dp} (P),Q_{1},\ldots,Q_{t_{2}})$. By
\eqref{lem:result1},
\[
\operatorname*{dp} (P,Q)=\operatorname*{dp}( M^{(1)})=\operatorname*{dp}%
(\operatorname*{dp} (P),Q_{1},\ldots,Q_{t_{2}})
\]

\item[Case (2):] $m_{2}=m_{1}-t_{1}$.

Let $U_{1}=\operatorname*{dp}(P_{1},\ldots,P_{t_{1}})$. Since $u_{i}$ is the
coefficient of $U_{1}$ in the term $x^{i}$, we have $u_{m_{1}-t_{1}%
+1}={\operatorname*{pcdp}}(P)$. Since $m_{2}=m_{1}-t_{1}$, $b_{j(m-t_{1}%
+1)}^{*}=0$ for $j=1,\ldots,t_{2}$, Equation \eqref{lem:result1} can be
simplified into the following:
\begin{align*}
\operatorname*{dp}(M^{(1)})=\operatorname*{dp}\left[
\setlength{\arraycolsep}{0.5pt}
\begin{array}
[c]{cccc}%
u_{m_{1}-t_{1}+1} & u_{m_{1}-t_{1}} & \cdots & u_{0}\\
& b_{1(m_{1}-t_{1})}^{*} & \cdots & b_{10}^{*}\\
& \vdots &  & \vdots\\
& b_{t_{2}(m_{1}-t_{1})}^{*} & \cdots & b_{t_{2}0}^{*}%
\end{array}
\right]  =u_{m_{1}-t_{1}+1}\cdot\operatorname*{dp}(Q)=\,{\operatorname*{pcdp}}
(P)\cdot\operatorname*{dp}(Q)
\end{align*}

\item[Case (3)] $m_{2}< m_{1}-t_{1}\wedge t_{2}= 1$.

Since $t_{2}=1$ and $m_{2}< m_{1}-t_{1}$, by \eqref{lem:result1},
\begin{align}
\label{equ:006}\operatorname*{dp}(M^{(1)})=\operatorname*{dp}\left[
\begin{array}
[c]{cccc}%
u_{m_{1}-t_{1}+1} & u_{m_{1}-t_{1}} & \cdots & u_{0}\\
& b_{1(m_{1}-t_{1})}^{*} & \cdots & b_{10}^{*}%
\end{array}
\right]  =u_{m_{1}-t_{1}+1}\cdot\sum_{i=0}^{m_{1}-t_{1}} b_{1i}^{*}\cdot x^{i}%
\end{align}
where $b_{ij}^{*}:=0$ when $j>q_{i}$. Noting that $u_{m_{1}-t_{1}%
+1}={\operatorname*{pcdp}} (P)$ and $\sum_{i=0}^{m_{1}-t_{1}} b_{1i}^{*}\cdot
x^{i}=Q_{1}=\operatorname*{dp} (Q)$, we easily attain
\[
\operatorname*{dp}(M^{(1)})={\operatorname*{pcdp}} (P)\cdot\operatorname*{dp}%
(Q)
\]

\item[Case (4)] $m_{2}<m_{1}-t_{1} \wedge t_{2}\geq2$.

Let $m_{2}=m_{1}-t_{1}-k$. Hence $b_{j(m_{1}-t_{1}+1)}^{*}=\cdots
=b_{j(m_{1}-t_{1}-k+1)}^{*}=0$ for $j=1,\ldots, t_{2}$. It follows that
\begin{align*}
\operatorname*{dp}(M^{(1)})=\,  &  \operatorname*{dp}(N^{(3)})\\
=\,  &  \operatorname*{dp}\left[
\begin{array}
[c]{cccc}%
u_{m_{1}-t_{1}+1} & u_{m_{1}-t_{1}} & \cdots & u_{0}\\
b_{1(m_{1}-t_{1}+1)}^{*} & b_{1(m_{1}-t_{1})}^{*} & \cdots & b_{10}^{*}\\
\vdots & \vdots &  & \vdots\\
b_{t_{2}(m_{1}-t_{1}+1)}^{*} & b_{t_{2}(m_{1}-t_{1})}^{*} & \cdots &
b_{t_{2}0}^{*}%
\end{array}
\right] \\
=\,  &  \operatorname*{dp}\left[
\begin{array}
[c]{ccccccc}%
u_{m_{1}-t_{1}+1} & u_{m_{1}-t_{1}} & \cdots & u_{m_{1}-t_{1}-k+1} &
u_{m_{1}-t_{1}-k} & \cdots & u_{0}\\
0 & 0 & \cdots & 0 & b_{1(m_{1}-t_{1}-k)}^{*} & \cdots & b_{10}^{*}\\
\vdots & \vdots &  & \vdots & \vdots &  & \vdots\\
0 & 0 & \cdots & 0 & b_{t_{2}(m_{1}-t_{1}-k)}^{*} & \cdots & b_{t_{2}0}^{*}%
\end{array}
\right] \\
=\,  &  u_{m_{1}-t_{1}+1}\cdot\operatorname*{dp}\left[
\begin{array}
[c]{ccccccccc}%
0 & \cdots & 0 & b_{1(m_{1}-t_{1}-k)}^{*} & \cdots & b_{10}^{*} &  &  & \\
\vdots &  & \vdots & \vdots &  & \vdots &  &  & \\
0 & \cdots & 0 & b_{t_{2}(m_{1}-t_{1}-k)}^{*} & \cdots & b_{t_{2}0}^{*} &  &
&
\end{array}
\right] \\
=\,  &  0
\end{align*}

\end{itemize}

\noindent The proof is completed.
\end{proof}

\subsection{Induction base (i.e. $k=1$)}

\label{subsec:base}

Note that when $k=1$, $v$ and $\epsilon$ can be explicitly written as
\begin{align}\label{equ:v:denpen:i}
v=\left\{
\begin{array}[c]{ll}
            (k,\ldots,k)+e_{0}=(k,\ldots,k)      & \text{if}\ \ i=0; \\
        (k,\ldots,k)+e_{i}=(k,\ldots,k,k+1,k\ldots,k) & \text{else}
\end{array}
\right.
\end{align}
and
\begin{align}\label{equ:epsilon:denpen:i}
\epsilon=\left\{
\begin{array}[c]{ll}
        |v+e_{0}|+k-2=(nk)+k-2=(n+1)k-2 & \text{if}\ \ i=0; \\
        |v+e_{i}|+k-2=(nk+2)+k-2=(n+1)k & \text{else}
\end{array}
\right.
\end{align}
respectively,  depending on whether $i$ is zero or not. For the sake of simplicity, let $k^{\prime}=(k,\ldots,k)$. By using \eqref{equ:v:denpen:i} and \eqref{equ:epsilon:denpen:i}, we can  provide an equivalent theorem of Main Result (Theorem
\ref{thm:main}) for $k=1$, which facilitates the proof.

\begin{theorem}
[Main Result for $k=1$]\label{thm:k=1} Let $w_{0}\in\mathbb{N}^{n}$ and
$i\in\left[  0,\ldots,n \right]  $ be such that $w_{0}+1^{\prime}+e_{i}\in
P(d_{0},n)$. Then we have
\begin{numcases}{\centering\hspace{-25mm}}\label{equ:base_step}
r_{w_{0}}^{n-1}(F) R_{w_{0}+1^{\prime}}(F) = R_{1^{\prime}}(R_{w_{0}+e_{0}}(F),R_{w_{0}+e_{1}}(F),\ldots,R_{w_{0}+e_{n}}(F))  \label{k=1:conclusion1}\\
r_{w_{0}}^{n+1}(F) R_{w_{0}+1^{\prime}+e_{i}}(F) = R_{1^{\prime}+e_{i}}(R_{w_{0}+e_{0}}(F),R_{w_{0}+e_{1}}(F),\ldots,R_{w_{0}+e_{n}}(F)) \ \ \ \text{if}\ i>0\label{k=1:conclusion2}
\end{numcases}

\end{theorem}

Here we give a sketch of the main idea adopted in the proof of induction base.
We start with a subresultant of the input polynomials with higher order (i.e.,
$w_{0}+1^{\prime}+e_{i}$). First, we use the multi-linearity of determinant
polynomial to rewrite it into a determinant polynomial of a polynomial set
consisting of the given polynomials and the subresultants with lower orders.
The tool we adopt in the rewriting process is the observation that every
subresultant belongs to the ideal generated by the input polynomials. Then we
employ the notable properties of determinant polynomials discovered in
Subsection \ref{ssec:detp_properties} to further simplify the determinant
polynomial obtained in the previous step, which will produce the result we want.

\subsubsection{Proof of Eq. \eqref{k=1:conclusion1}}

\begin{proof}
[Proof of Eq. \eqref{k=1:conclusion1}] Note that
\begin{align*}
R_{1^{\prime}}(R_{w_{0}+e_{0}},R_{w_{0}+e_{1}},\ldots,R_{w_{0}+e_{n}})
={\operatorname*{dp}}\ (R_{w_{0}+e_{1}},\ldots,R_{w_{0}+e_{n}})
\end{align*}
Let $\delta=w_{0}=(\delta_{1},\ldots,\delta_{n})$. Thus we only need to show
\begin{align*}
r_{\delta}^{n-1} \ R_{\delta+1^{\prime}} ={\operatorname*{dp}}\ (R_{\delta
+e_{1}},\ldots,R_{\delta+e_{n}})
\end{align*}
Let $G_{k,j}=x^{j}F_{k}$ and $H_{k,j}=(G_{k,j},\ldots,G_{k,0}).$ Then we have
\begin{align*}
R_{\delta+1^{\prime}}\ =\  &  \operatorname*{dp}(x^{\delta_{0}}F_{0}%
,\ldots,x^{0}F_{0},\ \ldots,\ x^{\delta_{n}}F_{n},\ldots,x^{0}F_{n})\\
=\  &  \operatorname*{dp}(H_{0,\delta_{0}},\ \ldots,\ H_{n,\delta_{n}})
\end{align*}
and for $i\ge1$,
\begin{align*}
R_{\delta+e_{i}}=  &  \left\{
\begin{array}
[c]{rl}%
{{\operatorname*{dp}}}(G_{0,\delta_{0}},H_{0,\delta_{0}-1},\ldots
,H_{i-1,\delta_{i-1}-1},H_{i,\delta_{i}},H_{i+1,\delta_{i+1}-1},\ldots
,H_{n,\delta_{n}-1}) & \text{if}\ \ d_{i}+\delta_{i}=d_{0}+\delta_{0};\\
{{\operatorname*{dp}}}(\ \ \ \ \ \ \ \ H_{0,\delta_{0}-1},\ldots
,H_{i-1,\delta_{i-1}-1},H_{i,\delta_{i}},H_{i+1,\delta_{i+1}-1},\ldots
,H_{n,\delta_{n}-1}) & \text{otherwise}%
\end{array}
\right.
\end{align*}
By Proposition \ref{prop:ci} and Corollary \ref{cor:dp}, $R_{\delta+e_{i}}$
can be written as the linear combination of $G_{i,j}$'s, i.e.,
\begin{align}
\label{equ:coeff_S+ei}R_{\delta+e_{i}}\ =\ \sum_{\substack{0\le t\le n}%
}^{}\sum_{j=0}^{\delta_{t}-1}c_{tj}^{(i)}G_{t,j}+c_{i\delta_{i}}%
^{(i)}G_{i,\delta_{i}}+\left\{
\begin{array}
[c]{ll}%
c_{0\delta_{0}}^{(0)}G_{0,\delta_{0}} & \text{if}\ \ d_{i}+\delta_{i}%
=d_{0}+\delta_{0};\\
0 & \text{otherwise}%
\end{array}
\right.
\end{align}
where
\begin{align}
\label{equ:015}c_{i\delta_{i}}^{(i)}=\left\{
\begin{array}
[c]{ll}%
(-1)^{\sigma_{i}+1}{\operatorname*{pcdp}}(G_{0,\delta_{0}},H_{0,\delta_{0}%
-1},\ldots,H_{n,\delta_{n}-1}) & \text{if}\ \ d_{i}+\delta_{i}=d_{0}%
+\delta_{0};\\
(-1)^{\sigma_{i}+1}{\operatorname*{pcdp}}(\ \ \ \ \ \ \ \ H_{0,\delta_{0}%
-1},\ldots,H_{n,\delta_{n}-1}) & \text{otherwise}%
\end{array}
\right.
\end{align}
and $\sigma_{i}=\delta_{i}+\cdots+\delta_{n}$. From the definition of
subresultant, we have
\begin{align*}
r_{\delta}={\operatorname*{pcdp}}(H_{0,\delta_{0}-1},\ldots,H_{n,\delta_{n}%
-1})
\end{align*}
Since $\deg G_{0,\delta_{0}}>\deg G_{i,j}$ for $i\ge0$ and $0\le j\le
\delta_{i}-1$, we simplify the first case of \eqref{equ:015} and obtain
\begin{align}
\label{equ:016}{\operatorname*{pcdp}}(G_{0,\delta_{0}},H_{0,\delta_{0}%
-1},\ldots,H_{n,\delta_{n}-1})\ =\ {\operatorname*{pcdp}}(G_{0,\delta_{0}%
})\cdot{\operatorname*{pcdp}}(H_{0,\delta_{0}-1},\ldots,H_{n,\delta_{n}-1})
=a_{0d_{0}}\cdot r_{\delta}=r_{\delta}%
\end{align}
Then substituting \eqref{equ:016} into the \eqref{equ:015}, we have
\begin{align}
c_{i\delta_{i}}^{(i)}=(-1)^{\sigma_{i}+1}r_{\delta}%
\end{align}
Hence
\begin{align*}
\prod_{i=1}^{n}c_{i\delta_{i}}^{(i)} =\prod_{i=1}^{n}(-1)^{\sigma_{i}+1}
r_{\delta} =r_{\delta}^{n}\prod_{i=1}^{n}(-1)^{\sigma_{i}+1}%
\end{align*}

\bigskip

\noindent We introduce the short-hands $\psi=\prod_{i=1}^{n}(-1)^{\sigma
_{i}+1}$ and $\rho=r_{\delta}^{n}$. Then $\prod_{i=1}^{n}c_{i\delta_{i}}%
^{(i)}=\psi\cdot\rho$. Now we consider the product of $\psi$, $\rho$ and
$R_{\delta+1^{\prime}}$ and carry out the following simplification:
\begin{align*}
&  \psi\ \rho\ R_{\delta+1^{\prime}}\\
=\  &  \prod_{i=1}^{n}c_{i\delta_{i}}^{(i)}{\operatorname*{dp}} (H_{0,\delta
_{0}},H_{1,\delta_{1}},\ldots,H_{n,\delta_{n}})\\
=\  &  \prod_{i=1}^{n}c_{i\delta_{i}}^{(i)}{\operatorname*{dp}}(H_{0,\delta
_{0}},\underset{H_{1,\delta_{1}}}{\underbrace{G_{1,\delta_{1}},H_{1,\delta
_{1}-1}}},\ldots,\underset{H_{n,\delta_{n}}}{\underbrace{G_{n,\delta_{n}%
},H_{n,\delta_{n}-1}}})\\
=\  &  \psi\ \prod_{i=1}^{n}c_{i\delta_{i}}^{(i)}{\operatorname*{dp}}
(H_{0,\delta_{0}},H_{1,\delta_{1}-1},\ldots,H_{n,\delta_{n}-1},G_{1,\delta
_{1}},\ldots,G_{n,\delta_{n}})\ \ \ \,\text{by reordering the arguments
of}\ {\operatorname*{dp}}\\
=\  &  \psi\ {\operatorname*{dp}} (H_{0,\delta_{0}},H_{1,\delta_{1}-1}%
,\ldots,H_{n,\delta_{n}-1},c_{1\delta_{1}}^{(1)}G_{1,\delta_{1}}%
,\ldots,c_{n\delta_{n}}^{(n)}G_{n,\delta_{n}}) \ \ \ \ \text{by\ pushing}%
\ c_{i\delta_{i}}^{(i)}\ \text{into} \ {\operatorname*{dp}}\\
=\  &  \psi\ {\operatorname*{dp}}(H_{0,\delta_{0}},H_{1,\delta_{1}-1}%
,\ldots,H_{n,\delta_{n}-1},R_{\delta+e_{1}},\ldots,R_{\delta+e_{n}})
\ \ \ \ \text{using}\ \text{}\ \eqref{equ:coeff_S+ei}\\
=\  &  \psi\ {\operatorname*{dp}}(\underset{H_{0,\delta_{0}}%
}{\underbrace{G_{0,\delta_{0}},H_{0,\delta_{0}-1}}},H_{1,\delta_{1}-1}%
,\ldots,H_{n,\delta_{n}-1},R_{\delta+e_{1}},\ldots,R_{\delta+e_{n}})\\
=\  &  \psi\ a_{0d_{0}}\ {\operatorname*{dp}}(H_{0,\delta_{0}-1}%
,H_{1,\delta_{1}-1},\ldots,H_{n,\delta_{n}-1},R_{\delta+e_{1}},\ldots
,R_{\delta+e_{n}}) \ \ \ \ \text{by}\ \text{Lemma\ \ref{lemma:002}-(2)}\\
=\  &  \psi\ {\operatorname*{dp}}(H_{0,\delta_{0}-1},H_{1,\delta_{1}-1}%
,\ldots,H_{n,\delta_{n}-1},R_{\delta+e_{1}},\ldots,R_{\delta+e_{n}})
\ \ \ \ \text{since}\ a_{0d_{0}}=1\
\end{align*}
The cancellation of $\psi$ from the first and last expressions yields
\begin{align*}
\rho\ R_{\delta+1^{\prime}} \ =\ {\operatorname*{dp}}(H_{0,\delta_{0}%
-1},H_{1,\delta_{1}-1},\ldots,H_{n,\delta_{n}-1},R_{\delta+e_{1}}%
,\ldots,R_{\delta+e_{n}})
\end{align*}
Partitioning the polynomial matrix in the right-hand side of the above
equation, we have $\rho R_{\delta+1^{\prime}} =\operatorname*{dp}(B_{1},B_{2})
$ where
\begin{align*}
B_{1}=(H_{0,\delta_{0}-1},H_{1,\delta_{1}-1},\ldots,H_{n,\delta_{n}%
-1}),\ \ B_{2}=(R_{\delta+e_{1}},\ldots,R_{\delta+e_{n}})
\end{align*}
Denote $\max(\deg B_{1})$, $\max(\deg B_{2})$ and the number of polynomials in
$B_{1}$ by $m_{1}$, $m_{2}$ and $\#B_{1}$, respectively. Then we have
$m_{1}=d_{0}+\delta_{0}-1$, $m_{2}=d_{0}-|\delta|-1$ and $\#B_{1}=\delta
_{0}+|\delta|$. It follows that
\[
m_{2}=m_{1}-\#B_{1}
\]
By Lemma \ref{lemma:002}-(2), we have
\begin{align}
\label{equ:017}r_{\delta}^{n}\ R_{\delta+1^{\prime}} \ =\rho R_{\delta
+1^{\prime}}=\ {\operatorname*{pcdp}}(B_{1})\cdot\operatorname*{dp}(B_{2})
\end{align}
It is easy to see that
\begin{align*}
\operatorname*{dp}(B_{1}) \ =\ \operatorname*{dp}(H_{0,\delta_{0}%
-1},H_{1,\delta_{1}-1},\ldots,H_{n,\delta_{n}-1}) \ =\ R_{\delta}%
\end{align*}
which immediately implies that ${\operatorname*{pcdp}}(B_{1})=r_{\delta}$.
Substituting it into \eqref{equ:017} yields
\begin{align*}
r_{\delta}^{n} R_{\delta+1^{\prime}} \ =\ {\operatorname*{pcdp}}(B_{1}%
)\cdot\operatorname*{dp}(B_{2}) \ =\ r_{\delta}\cdot\operatorname*{dp}%
(R_{\delta+e_{1}},\ldots,R_{\delta+e_{n}})
\end{align*}
It follows that
\begin{align*}
r_{\delta}^{n-1} R_{\delta+1^{\prime}} \ =\ \operatorname*{dp}(R_{\delta
+e_{1}},\ldots,R_{\delta+e_{n}})
\end{align*}
Eq. \eqref{k=1:conclusion1} is proved.
\end{proof}

\subsubsection{Proof of Eq. \eqref{k=1:conclusion2}}

\begin{proof}
[Proof of Eq. \eqref{k=1:conclusion2}] Without loss of generality, we only
prove the case when $i=1$, that is,
\[
R_{1^{\prime}+e_{1}}(R_{w_{0}+e_{0}},R_{w_{0}+e_{1}},\ldots,R_{w_{0}+e_{n}})
\ =\ {\operatorname*{dp}}(R_{w_{0}+e_{0}},xR_{w_{0}+e_{1}},R_{w_{0}+e_{1}%
},\ldots,R_{w_{0}+e_{n}})
\]
The other cases of $i>0$ can be proved with the same procedure.

\medskip

Let $\delta=w_{0}=(\delta_{1},\ldots,\delta_{n})$. Thus we only need to show
that
\begin{align*}
r_{\delta}^{n+1} R_{\delta+1^{\prime}+e_{1}} ={\operatorname*{dp}}%
(R_{\delta+e_{0}},xR_{\delta+e_{1}},R_{\delta+e_{1}},\ldots,R_{\delta+e_{n}})
\end{align*}
Let $G_{k,j}=x^{j}F_{k}$ and $H_{k,j}=(G_{k,j},\ldots,G_{k,0})$. Then we have
\begin{align*}
R_{\delta+1^{\prime}+e_{1}} =\left\{
\begin{array}
[c]{rl}%
{{\operatorname*{dp}}}(G_{0,\delta_{0}+1},H_{0,\delta_{0}},H_{1,\delta_{1}%
+1},H_{2,\delta_{2}},\ldots,H_{n,\delta_{n}}) & \text{if}\ \ d_{1}+\delta
_{1}=d_{0}+\delta_{0};\\
{{\operatorname*{dp}}}(\ \ \ \ \ \ \ \ \ \ \ H_{0,\delta_{0}},H_{1,\delta
_{1}+1},H_{2,\delta_{2}},\ldots,H_{n,\delta_{n}}) & \text{otherwise}%
\end{array}
\right.
\end{align*}
Recall \eqref{equ:coeff_S+ei}, that is, for $i\ge1$,
\begin{align}
\label{equ:027}R_{\delta+e_{i}}\ =\ \sum_{\substack{0\le t\le n}}^{}\sum
_{j=0}^{\delta_{t}-1}c_{tj}^{(i)}G_{t,j}+c_{i\delta_{i}}^{(i)}G_{i,\delta_{i}%
}+\left\{
\begin{array}
[c]{ll}%
c_{0\delta_{0}}^{(0)}G_{0,\delta_{0}} & \text{if}\ \ d_{i}+\delta_{i}%
=d_{0}+\delta_{0};\\
0 & \text{otherwise}%
\end{array}
\right.
\end{align}
where
\begin{align}
\label{equ:019}c_{i\delta_{i}}^{(i)}=(-1)^{\sigma_{i}+1}r_{\delta}%
\end{align}
and $\sigma_{i}=\delta_{i}+\cdots+\delta_{n}$.

By \eqref{equ:019}, we have
\begin{align*}
c_{1\delta_{1}}^{(1)}\prod_{i=1}^{n}c_{i\delta_{i}}^{(i)} =(-1)^{\sigma_{1}%
+1}r_{\delta}\cdot\prod_{i=1}^{n}(-1)^{\sigma_{i}+1}r_{\delta} =(-1)^{\sigma
_{1}+1}\cdot r_{\delta}^{n+1}\cdot\prod_{i=1}^{n}(-1)^{\sigma_{i}+1}%
\end{align*}

Again we introduce the short-hands $\psi=(-1)^{\sigma_{1}+1}\prod_{i=1}%
^{n}(-1)^{\sigma_{i}+1}$ and $\rho=r_{\delta}^{n+1}$. Then $c_{1\delta_{1}%
}^{(1)}\prod_{i=1}^{n}c_{i\delta_{i}}^{(i)}=\psi\cdot\rho$. Now we consider
the following two cases for the product of $\psi$, $\rho$ and $R_{\delta
+1^{\prime}+e_{1}}$, depending on whether $d_{1}+\delta_{1}=d_{0}+\delta_{0}$
holds or not. \medskip

\textbf{Case (1)}: $d_{1}+\delta_{1}=d_{0}+\delta_{0}$. In this case,
\begin{align*}
&  \psi\ \rho\ R_{\delta+1^{\prime}+e_{1}}\\
\ =\  &  c_{1\delta_{1}}^{(1)}\prod_{i=1}^{n}c_{i\delta_{i}}^{(i)}%
\ {\operatorname*{dp}}(H_{0,\delta_{0}+1},H_{1,\delta_{1}+1},H_{2,\delta_{2}%
},\ldots,H_{n,\delta_{n}})\\
\ =\  &  c_{1\delta_{1}}^{(1)}\prod_{i=1}^{n}c_{i\delta_{i}}^{(i)}%
\ {\operatorname*{dp}}(H_{0,\delta_{0}+1},\underset{H_{1,\delta_{1}%
+1}}{\underbrace{G_{1,\delta_{1}+1},G_{1,\delta_{1}},H_{1,\delta_{1}-1}}%
},\underset{H_{2,\delta_{2}}}{\underbrace{G_{2,\delta_{2}},H_{2,\delta_{2}-1}%
}},\ldots,\underset{H_{n,\delta_{n}}}{\underbrace{G_{n,\delta_{n}}%
,H_{n,\delta_{n}-1}}})\\
\ =\  &  \psi\ c_{1\delta_{1}}^{(1)}\prod_{i=1}^{n}c_{i\delta_{i}}%
^{(i)}\ {\operatorname*{dp}}(H_{0,\delta_{0}+1},H_{1,\delta_{1}-1}%
,H_{2,\delta_{2}-1},\ldots,H_{n,\delta_{n}-1},G_{1,\delta_{1}+1},
G_{1,\delta_{1}},G_{2,\delta_{2}},\ldots,G_{n,\delta_{n}})\\
&  \text{(by reordering the arguments of}\ {\operatorname*{dp})}\\
\ =\  &  \psi\ {\operatorname*{dp}}(H_{0,\delta_{0}+1},H_{1,\delta_{1}%
-1},H_{2,\delta_{2}-1},\ldots,H_{n,\delta_{n}-1},c_{1\delta_{1}}%
^{(1)}G_{1,\delta_{1}+1}, c_{1\delta_{1}}^{(1)}G_{1,\delta_{1}},c_{2\delta
_{2}}^{(2)}G_{2,\delta_{2}},\ldots,c_{n\delta_{n}}^{(n)}G_{n,\delta_{n}})\\
&  \text{(by\ pushing}\ c_{i\delta_{i}}^{(i)}\ \text{into}
\ {\operatorname*{dp)}}\\
\ =\  &  \psi\ {\operatorname*{dp}} (H_{0,\delta_{0}+1},H_{1,\delta_{1}%
-1},H_{2,\delta_{2}-1},\ldots,H_{n,\delta_{n}-1},xR_{\delta+e_{1}}%
,R_{\delta+e_{1}},R_{\delta+e_{2}},\ldots,R_{\delta+e_{n}})\ \ \ \ \text{by}%
\ \eqref{equ:027}\\
\ =\  &  \psi\ {\operatorname*{dp}} (\underset{H_{0,\delta_{0}+1}%
}{\underbrace{G_{0,\delta_{0}+1},G_{0,\delta_{0}},H_{0,\delta_{0}-1}}%
},H_{1,\delta_{1}-1},H_{2,\delta_{2}-1},\ldots,H_{n,\delta_{n}-1}%
,xR_{\delta+e_{1}},R_{\delta+e_{1}},R_{\delta+e_{2}},\ldots,R_{\delta+e_{n}%
})\\
\ =\  &  \psi\ a_{0d_{0}}^{2} {\operatorname*{dp}} (H_{0,\delta_{0}%
-1},H_{1,\delta_{1}-1},H_{2,\delta_{2}-1},\ldots,H_{n,\delta_{n}-1}%
,xR_{\delta+e_{1}},R_{\delta+e_{1}},R_{\delta+e_{2}},\ldots,R_{\delta+e_{n}%
})\ \ \ \ \text{by}\ \text{Lemma\ \ref{lemma:002}-(2)}\\
\ =\  &  \psi\ {\operatorname*{dp}} (H_{0,\delta_{0}-1},H_{1,\delta_{1}%
-1},H_{2,\delta_{2}-1},\ldots,H_{n,\delta_{n}-1},xR_{\delta+e_{1}}%
,R_{\delta+e_{1}},R_{\delta+e_{2}},\ldots,R_{\delta+e_{n}}%
)\ \ \ \ \text{since}\ a_{0d_{0}}=1
\end{align*}

\medskip

\textbf{Case (2)}: $d_{1}+\delta_{1}\neq d_{0}+\delta_{0}$.
\begin{align*}
&  \psi\ \rho\ R_{\delta+1^{\prime}+e_{1}}\\
\ =\  &  c_{1\delta_{1}}^{(1)}\prod_{i=1}^{n}c_{i\delta_{i}}^{(i)}%
{\operatorname*{dp}} (H_{0,\delta_{0}},H_{1,\delta_{1}+1},H_{2,\delta_{2}%
},\ldots,H_{n,\delta_{n}})\\
\ =\  &  c_{1\delta_{1}}^{(1)}\prod_{i=1}^{n}c_{i\delta_{i}}^{(i)}%
{\operatorname*{dp}} (H_{0,\delta_{0}},\underset{H_{1,\delta_{1}%
+1}}{\underbrace{G_{1,\delta_{1}+1},G_{1,\delta_{1}},H_{1,\delta_{1}-1}}%
},\underset{H_{2,\delta_{2}}}{\underbrace{G_{2,\delta_{2}},H_{2,\delta_{2}-1}%
}},\ldots,\underset{H_{n,\delta_{n}}}{\underbrace{G_{n,\delta_{n}}%
,H_{n,\delta_{n}-1}}})\\
\ =\  &  \psi\ c_{1\delta_{1}}^{(1)}\prod_{i=1}^{n}c_{i\delta_{i}}%
^{(i)}{\operatorname*{dp}}(H_{0,\delta_{0}},H_{1,\delta_{1}-1},H_{2,\delta
_{2}-1},\ldots,H_{n,\delta_{n}-1},G_{1,\delta_{1}+1},G_{1,\delta_{1}%
},G_{2,\delta_{2}},\ldots,G_{n,\delta_{n}})\\
&  \text{(by reordering the arguments of}\ {\operatorname*{dp}})\\
\ =\  &  \psi\ {\operatorname*{dp}}(H_{0,\delta_{0}},H_{1,\delta_{1}%
-1},H_{2,\delta_{2}-1},\ldots,H_{n,\delta_{n}-1},c_{1\delta_{1}}%
^{(1)}G_{1,\delta_{1}+1},c_{1\delta_{1}}^{(1)}G_{1,\delta_{1}},c_{2\delta_{2}%
}^{(2)}G_{2,\delta_{2}},\ldots,c_{n\delta_{n}}^{(n)}G_{n,\delta_{n}})\\
&  \text{(by\ pushing}\ c_{i\delta_{i}}^{(i)}\ \text{into}
\ {\operatorname*{dp}})\\
\ =\  &  \psi\ {\operatorname*{dp}}(H_{0,\delta_{0}},H_{1,\delta_{1}%
-1},H_{2,\delta_{2}-1},\ldots,H_{n,\delta_{n}-1},xR_{\delta+e_{1}}%
,R_{\delta+e_{1}},R_{\delta+e_{2}},\ldots,R_{\delta+e_{n}})\ \ \ \ \text{by}%
\ \eqref{equ:027}\\
\ =\  &  \psi\ {\operatorname*{dp}}(\underset{H_{0,\delta_{0}}%
}{\underbrace{G_{0,\delta_{0}},H_{0,\delta_{0}-1}}},H_{1,\delta_{1}%
-1},H_{2,\delta_{2}-1},\ldots,H_{n,\delta_{n}-1},xR_{\delta+e_{1}}%
,R_{\delta+e_{1}},R_{\delta+e_{2}},\ldots,R_{\delta+e_{n}})\\
\ =\  &  \psi\ a_{0d_{0}}\ {\operatorname*{dp}}(H_{0,\delta_{0}-1}%
,H_{1,\delta_{1}-1},H_{2,\delta_{2}-1},\ldots,H_{n,\delta_{n}-1}%
,xR_{\delta+e_{1}},R_{\delta+e_{1}},R_{\delta+e_{2}},\ldots,R_{\delta+e_{n}%
})\ \ \ \ \text{by}\ \text{Lemma\ \ref{lemma:002}-(2)}\\
\ =\  &  \psi\ {\operatorname*{dp}}(H_{0,\delta_{0}-1},H_{1,\delta_{1}%
-1},H_{2,\delta_{2}-1},\ldots,H_{n,\delta_{n}-1},xR_{\delta+e_{1}}%
,R_{\delta+e_{1}},R_{\delta+e_{2}},\ldots,R_{\delta+e_{n}}%
)\ \ \ \ \text{since}\ a_{0d_{0}}=1
\end{align*}
It can be seen that regardless of whether $d_{1}+\delta_{1}=d_{0}+\delta_{0}$
holds or not, the following equation is always true:
\begin{align*}
\psi\ \rho\ R_{\delta+1^{\prime}+e_{1}} \ =\ \psi\ {\operatorname*{dp}%
}(H_{0,\delta_{0}-1},H_{1,\delta_{1}-1},H_{2,\delta_{2}-1},\ldots
,H_{n,\delta_{n}-1},xR_{\delta+e_{1}},R_{\delta+e_{1}},R_{\delta+e_{2}}%
,\ldots,R_{\delta+e_{n}})
\end{align*}
The cancellation of $\psi$ in the left-hand side and right-hand side of above
expressions yields
\begin{align}
\label{equ:020}\rho\ R_{\delta+1^{\prime}+e_{1}} \ =\ {\operatorname*{dp}%
}(H_{0,\delta_{0}-1},H_{1,\delta_{1}-1},H_{2,\delta_{2}-1},\ldots
,H_{n,\delta_{n}-1},xR_{\delta+e_{1}},R_{\delta+e_{1}},R_{\delta+e_{2}}%
,\ldots,R_{\delta+e_{n}})
\end{align}
Partitioning the polynomial matrix in the right-hand side of \eqref{equ:020},
we have $\rho\ R_{\delta+1^{\prime}+e_{1}}={\operatorname*{dp}}(B_{1},B_{2}),
$ where
\begin{align*}
B_{1}\ =\ (H_{0,\delta_{0}-1},H_{1,\delta_{1}-1},H_{2,\delta_{2}-1}%
,\ldots,H_{n,\delta_{n}-1}), \ \ \ B_{2}\ =\ (xR_{\delta+e_{1}},R_{\delta
+e_{1}},R_{\delta+e_{2}},\ldots,R_{\delta+e_{n}})
\end{align*}
Denote $\max(\deg B_{1})$, $\max(\deg B_{2})$ and the number of polynomials in
$B_{1}$ by $m_{1}$, $m_{2}$ and $\#B_{1}$, respectively. Then we have
$m_{1}=d_{0}+\delta_{0}-1$, $m_{2}=d_{0}-|\delta|$ and $\#B_{1}=|\delta
|+\delta_{0}$. It follows that
\begin{align*}
m_{2}\ =\ m_{1}-\#B_{1}+1
\end{align*}
By Lemma \ref{lemma:002}-(1), we have
\begin{align}
\label{equ:021}r_{\delta}^{n+1}\ R_{\delta+1^{\prime}+e_{1}}\ =\rho
\ R_{\delta+1^{\prime}+e_{1}}=\ {\operatorname*{dp}}(\operatorname*{dp}%
(B_{1}),B_{2})
\end{align}
It easy to know
\begin{align}
\label{equ:022}\operatorname*{dp}(B_{1}) \ =\ \operatorname*{dp}%
(H_{0,\delta_{0}-1},H_{1,\delta_{1}-1},\ldots,H_{n,\delta_{n}-1})
\ =\ R_{\delta}%
\end{align}
The substitution of \eqref{equ:022} into \eqref{equ:021} yields%
\begin{align*}
r_{\delta}^{n+1}\ R_{\delta+1^{\prime}+e_{1}} \ =\  &  {\operatorname*{dp}}
(R_{\delta},xR_{\delta+e_{1}},R_{\delta+e_{1}},\ldots,R_{\delta+e_{n}})
\end{align*}
Eq. \eqref{k=1:conclusion2} is proved.
\end{proof}

\subsection{Inductive step (i.e. from $k=j$ to $k=j+1$)}

Similarly to the simplification as done in the case $k=1$ (i.e., Theorem \ref{thm:k=1}), we let $k^{\prime}=(k,\ldots,k)$. Again by using \eqref{equ:v:denpen:i} and \eqref{equ:epsilon:denpen:i}, we  provide an equivalent theorem of Main Result (Theorem \ref{thm:main}) for $k\geq2$, which facilitates the proof.

\begin{theorem}
[Main Result for $k\geq2$]\label{thm:k>=2} Assume Theorem \ref{thm:main} holds
for $v=j^{\prime}+e_{i}$ (i.e., $k=j$), that is,
\begin{align}
r_{w_{0}}^{(n+1)j-2} R_{w_{0}+j^{\prime}}  &  =R_{j^{\prime}}(R_{w_{0}+e_{0}%
},\ldots,R_{w_{0}^{\prime}+e_{n}})\label{equ:theorem:k=j_1}\\
r_{w_{0}}^{(n+1)j} R_{w_{0}+j^{\prime}+e_{i}}  &  =R_{j^{\prime}+e_{i}%
}(R_{w_{0}+e_{0}},\ldots,R_{w_{0}+e_{n}}) \ \ \text{for $i>0$}
\label{equ:theorem:k=j_2}%
\end{align}
where $n+1=\# F$. Then the theorem is true for $v=j^{\prime}+1^{\prime}+e_{i}$
(i.e., $k=j+1$), that is,
\begin{align}
r_{w_{0}}^{(n+1)(j+1)-2} R_{w_{0}+j^{\prime}+1^{\prime}}  &  =R_{j^{\prime
}+1^{\prime}}(R_{w_{0}+e_{0}},\ldots,R_{w_{0}+e_{n}}%
)\label{equ:theorem:k=j+1_1}\\
r_{w_{0}}^{(n+1)(j+1)} R_{w_{0}+j^{\prime}+1^{\prime}+e_{i}}  &
=R_{j^{\prime}+1^{\prime}+e_{i}}(R_{w_{0}+e_{0}}, \ldots,R_{w_{0}+e_{n}})
\ \ \text{for $i>0$} \label{equ:theorem:k=j+1_2}%
\end{align}

\end{theorem}

Here we give a sketch of the main idea adopted in the inductive proof. In the
inductive step, we first use the relationship proved in the induction base to
write the subresultant of the input polynomials with higher order (i.e.,
$w_{0}+j^{\prime}+1^{\prime}+e_{i}$) into a determinant polynomial of
subresultants of order $w_{0}+j^{\prime}+e_{\tau}$, where $\tau=0,\ldots,n$.
This is the most key step in the inductive proof. Then we use the same
techniques (i.e., multi-linearity of determinant polynomial and the membership
of subresultant in the ideal generated by the defining polynomials) as in the
induction base to simplify the nested subresultants in the right-hand side of
\eqref{equ:theorem:k=j+1_1} and \eqref{equ:theorem:k=j+1_2}. The only
difference from the induction base is that the defining polynomials of the
internal subresultant is, by induction, subresultants in the previous step.
Again we employ the properties of determinant polynomials in Subsection
\ref{ssec:detp_properties} to further simplify the determinant polynomial
obtained in the previous step, which will produce the result we want.

\label{subsec:inductive}

\subsubsection{Proof of Eq. \eqref{equ:theorem:k=j+1_1} from Eqs.
\eqref{equ:theorem:k=j_1}--\eqref{equ:theorem:k=j_2}}

The goal of this part is to prove
\begin{align*}
r_{w_{0}}^{(n+1)(j+1)-2} R_{w_{0}+j^{\prime}+1^{\prime}}=R_{j^{\prime
}+1^{\prime}}(R_{w_{0}+e_{0}},\ldots,R_{w_{0}+e_{n}})
\end{align*}
with \eqref{equ:theorem:k=j_1} and \eqref{equ:theorem:k=j_2} assumed to be true.

\begin{proof}
[Proof of Eq. \eqref{equ:theorem:k=j+1_1}] By Lemma \ref{thm:k=1}, we get
\begin{align*}
r_{w_{0}+j^{\prime}}^{n-1} \ R_{w_{0}+j^{\prime}+1^{\prime}}  &
\ =\ R_{1^{\prime}}(R_{w_{0}+j^{\prime}+e_{0}},\ldots,R_{w_{0}+j^{\prime
}+e_{n}})\\
&  \ =\ {\operatorname*{dp}}(R_{w_{0}+j^{\prime}+e_{1}},\ldots,R_{w_{0}%
+j^{\prime}+e_{n}})
\end{align*}
Let $\delta=w_{0}=(\delta_{1},\ldots,\delta_{n})$. Thus we only need to show
that
\begin{equation}
\label{equ:equivalent1}{\operatorname*{dp}}(R_{\delta+j^{\prime}+e_{1}}%
,\ldots,R_{\delta+j^{\prime}+e_{n}}) \ =\ r_{\delta}^{2-(n+1)(j+1)}
r_{\delta+j^{\prime}}^{n-1}\ R_{j^{\prime}+1}(R_{\delta+e_{0}},\ldots
,R_{\delta+e_{n}})
\end{equation}

By induction (i.e. \eqref{equ:theorem:k=j_2}), for $i>0$,
\begin{align}
\label{equ:uij_1}r_{\delta}^{(n+1)j}\ R_{\delta+j^{\prime}+e_{i}}%
=R_{j^{\prime}+e_{i}}(R_{\delta+e_{0}},\ldots,R_{\delta+e_{n}})
\end{align}
Let $\hat{G}_{i,s}=x^{s}R_{\delta+e_{i}}$ and $\hat{H}_{i,s}=(\hat{G}%
_{i,s},\ldots,\hat{G}_{i,0}$). Then we have
\begin{align*}
R_{j^{\prime}+e_{i}}(R_{\delta+e_{0}},\ldots,R_{\delta+e_{n}}) \ =\  &
\operatorname*{dp}(\hat{H}_{0,j-1},\ldots,\hat{H}_{i-1,j-1},\hat{H}_{i,j}%
,\hat{H}_{i+1,j-1},\ldots,\hat{H}_{n,j-1})
\end{align*}
for $i>0$. Thus
\begin{align*}
r_{\delta}^{(n+1)j}\ R_{\delta+j^{\prime}+e_{i}}  &  \ =\ \operatorname*{dp}%
(\hat{H}_{0,j-1},\ldots,\hat{H}_{i-1,j-1},\hat{H}_{i,j},\hat{H}_{i+1,j-1}%
,\ldots,\hat{H}_{n,j-1})
\end{align*}
By Corollary \ref{cor:dp}, we deduce that
\begin{align}
\label{equ:uij_2}r_{\delta}^{(n+1)j}\ R_{\delta+j^{\prime}+e_{i}}  &
\ =\ \sum_{t=0}^{n}\sum_{h=0}^{j-1}u_{th}^{(i)}\hat{G}_{t,h}+u_{ij}^{(i)}%
\hat{G}_{i,j}%
\end{align}
for $i>0$. Note that
\begin{align}
u_{ij}^{(i)}  &  =(-1)^{(n+1-i)j}\ {\operatorname*{pcdp}}%
(\ \ \ \ \ \ \ \ \ \ \hat{H}_{0,j-1}, \ldots, \hat{H}_{i-1,j-1},\hat
{H}_{i,j-1},\hat{H}_{i+1,j-1},\ldots,\hat{H}_{n,j-1})\nonumber\\
&  =(-1)^{(n+1-i)j}\ {\operatorname*{pcdp}}(\underset{\hat{H}_{0,j-1}%
}{\underbrace{\hat{G}_{0,j-1},\hat{H}_{0,j-2}}},\ldots,\hat{H}_{i-1,j-1}%
,\hat{H}_{i,j-1},\hat{H}_{i+1,j-1},\ldots,\hat{H}_{n,j-1})\nonumber\\
&  =(-1)^{(n+1-i)j}\ {\operatorname*{pcdp}}(\hat{G}_{0,j-1})\ \cdot
{\operatorname*{pcdp}}(\hat{H}_{0,j-2},\hat{H}_{1,j-1},\ldots,\hat
{H}_{i-1,j-1},\hat{H}_{i,j-1},\hat{H}_{i+1,j-1},\ldots,\hat{H}_{n,j-1}%
)\nonumber\\
&  =(-1)^{(n+1-i)j}\ r_{\delta}\cdot r_{j^{\prime}}(R_{\delta+e_{0}}%
,\ldots,R_{\delta+e_{n}}) \label{equ:010}%
\end{align}
In addition, by the assumption \eqref{equ:theorem:k=j_1}, we have
\begin{align}
r_{\delta}^{(n+1)j-2} r_{\delta+j^{\prime}}  &  =r_{j^{\prime}}(R_{\delta
+e_{0}},\ldots,R_{\delta+e_{n}}) \label{equ:009}%
\end{align}
The substitution of \eqref{equ:009} into \eqref{equ:010} yields
\begin{align}
\label{equ:uij_3}u_{ij}^{(i)} \ =\ (-1)^{(n+1-i)j}\ r_{\delta} \cdot
r_{\delta}^{(n+1)j-2}\ r_{\delta+j^{\prime}} \ =\ (-1)^{(n+1-i)j}\ r_{\delta
}^{(n+1)j-1}\ r_{\delta+j^{\prime}}%
\end{align}
By \eqref{equ:uij_3}, we have
\begin{align*}
\prod_{i=1}^{n}u_{ij}^{(i)} \ =\ \prod_{i=1}^{n}\left(  (-1)^{(n+1-i)j}
r_{\delta}^{(n+1)j-1} r_{\delta+j^{\prime}}\right)  \ =\ \left(  \prod
_{i=1}^{n}(-1)^{(n+1-i)j}\right)  \cdot(r_{\delta}^{(n+1)j-1} r_{\delta
+j^{\prime}})^{n}%
\end{align*}
Now we introduce the short-hands $\psi_{j} =\prod_{i=1}^{n}(-1)^{(n+1-i)j}$
and $\rho_{j} =r_{\delta}^{(n+1)j}$. Then
\[
\prod_{i=1}^{n}u_{ij}^{(i)}\ =\ \psi_{j}\cdot(\rho_{j}r_{\delta}^{-1}
r_{\delta+j^{\prime}})^{n}%
\]
Next we consider the product of $\psi_{j}$, $(\rho_{j}r_{\delta}^{-1}
r_{\delta+j^{\prime}})^{n}$ and $R_{j^{\prime}+1}(R_{\delta+e_{0}}%
,\ldots,R_{\delta+e_{n}})$ and carry out the following simplification:
\begin{align*}
&  \psi_{j}\ \ (\rho_{j}r_{\delta}^{-1} r_{\delta+j^{\prime}})^{n}%
\ R_{j^{\prime}+1}(R_{\delta+e_{0}},\ldots,R_{\delta+e_{n}})\\
\ =\  &  \prod_{i=1}^{n}u_{ij}^{(i)}\ \operatorname*{dp}(\hat{H}_{0,j-1}%
,\hat{H}_{1,j},\ldots,\hat{H}_{n,j})\\
\ =\  &  \prod_{i=1}^{n}u_{ij}^{(i)}\ \operatorname*{dp}(\hat{H}%
_{0,j-1},\underset{\hat{H}_{1,j}}{\underbrace{\hat{G}_{1,j},\hat{H}_{1,j-1}}%
},\ldots,\underset{\hat{H}_{n,j}}{\underbrace{\hat{G}_{n,j},\hat{H}_{n,j-1}}%
})\\
\ =\  &  \psi_{j}\ \prod_{i=1}^{n}u_{ij}^{(i)}\ \operatorname*{dp}(\hat
{H}_{0,j-1},\hat{H}_{1,j-1},\ldots,\hat{H}_{n,j-1},\hat{G}_{1,j}\ldots,\hat
{G}_{n,j})\ \ \text{by reordering the arguments of}\ {\operatorname*{dp}}\\
\ =\  &  \psi_{j}\ {\operatorname*{dp}}(\hat{H}_{0,j-1},\hat{H}_{1,j-1}%
,\ldots,\hat{H}_{n,j-1},u_{1j}^{(1)}\hat{G}_{1,j},\ldots,u_{nj}^{(n)}\hat
{G}_{n,j})\ \ \text{by\ pushing}\ u_{ij}^{(i)}\ \text{into}
\ {\operatorname*{dp}}\\
\ =\  &  \psi_{j}\ {\operatorname*{dp}}(\hat{H}_{0,j-1},\hat{H}_{1,j-1}%
,\ldots,\hat{H}_{n,j-1},\rho_{j}\ R_{\delta+j^{\prime}+e_{1}},\ldots,\rho
_{j}\ R_{\delta+j^{\prime}+e_{n}}) \ \ \text{by}\ \eqref{equ:uij_2}\\
\ =\  &  \psi_{j}\ \rho_{j}^{n}\ {\operatorname*{dp}}(\hat{H}_{0,j-1},\hat
{H}_{1,j-1},\ldots,\hat{H}_{n,j-1},R_{\delta+j^{\prime}+e_{1}},\ldots,
R_{\delta+j^{\prime}+e_{n}}) \ \ \text{by\ pulling\ out }\rho_{j}%
\ \text{out\ of }{\operatorname*{dp}}%
\end{align*}
The cancellation of $\psi_{j}\rho_{j}^{n}$ from the first and last expressions
yields
\begin{align}
r_{\delta+j^{\prime}}^{n}\ r_{\delta}^{-n}\ R_{j^{\prime}+1}(R_{\delta+e_{0}%
},\ldots,R_{\delta+e_{n}}) \ =\ {\operatorname*{dp}}(\hat{H}_{0,j-1},\hat
{H}_{1,j-1},\ldots,\hat{H}_{n,j-1},R_{\delta+j^{\prime}+e_{1}},\ldots,
R_{\delta+j^{\prime}+e_{n}}) \label{eq:partitioned_matrix}%
\end{align}
Partitioning the polynomial matrix in the right-hand side of
\eqref{eq:partitioned_matrix}, we have
\begin{align}
\label{equ:003}r_{\delta+j^{\prime}}^{n}\ r_{\delta}^{-n}\ R_{j^{\prime
}+1^{\prime}}(R_{\delta+e_{0}},\ldots,R_{\delta+e_{n}}) ={\operatorname*{dp}%
}(B_{1},B_{2})
\end{align}
where
\begin{align*}
B_{1}=(\hat{H}_{0,j-1},\hat{H}_{1,j-1},\ldots,\hat{H}_{n,j-1}),\qquad
B_{2}=(R_{\delta+j^{\prime}+e_{1}},\ldots, R_{\delta+j^{\prime}+e_{n}})
\end{align*}
Denote $\max(\deg B_{1})$, $\max(\deg B_{2})$ and the number of polynomials in
$B_{1}$ by $m_{1}$, $m_{2}$ and $\#B_{1}$, respectively. Then we have
$m_{1}=d_{0}-|\delta|+j-1$, $m_{2}=d_{0}-|\delta|-nj-1$ and $\#B_{1}=(n+1)j$.
It follows that
\begin{align*}
m_{2}=m_{1}-\#B_{1}%
\end{align*}
By Lemma \ref{lemma:002}-(2), we derive the following
\begin{align}
\label{equ:023}{\operatorname*{dp}}(B_{1},B_{2}) = {\operatorname*{pcdp}%
}(B_{1})\cdot{\operatorname*{dp}}(B_{2})
\end{align}
In order to have an explicit expression of ${\operatorname*{pcdp}}(B_{1})$, we
carry out the following deduction:
\begin{align*}
\operatorname*{dp}(B_{1})\ =\  &  \operatorname*{dp}(\ \ \ \ \ \ \ \ \ \ \hat
{H}_{0,j-1},\hat{H}_{1,j-1},\ldots,\hat{H}_{n,j-1})\\
\ =\  &  \operatorname*{dp}(\underset{\hat{H}_{0,j-1}}{\underbrace{\hat
{G}_{0,j-1},\hat{H}_{0,j-2}}},\hat{H}_{1,j-1},\ldots,\hat{H}_{n,j-1})\\
\ =\  &  { \operatorname*{pcdp}}(\hat{G}_{0,j-1})\cdot\operatorname*{dp}%
(\hat{H}_{0,j- 2},\hat{H}_{1,j-1},\ldots,\hat{H}_{n,j-1}%
)\ \ \ \ \ \text{by\ Lemma\ \ref{lemma:002}-(2)}\\
\ =\  &  {\operatorname*{pcdp}}(\hat{G}_{0,j-1})\cdot R_{j^{\prime}}%
(R_{\delta},\ldots,R_{\delta^{\prime}+e_{n}})\\
\ =\  &  r_{\delta}\cdot r_{ \delta}^{(n+1)j-2} R_{\delta+j^{\prime}}
\ \ \ \ \ \text{by\ the\ assumption}\ \eqref{equ:theorem:k=j_1}\\
\ =\  &  r_{\delta}^{(n+1)j -1} R_{\delta+j^{\prime}}%
\end{align*}
which implies ${\operatorname*{pcdp}}(B_{1})=r_{\delta}^{(n+1)j-1}
r_{\delta+j^{\prime}}$. After substituting it into \eqref{equ:023}, we have
\begin{align}
\label{equ:025}\operatorname*{dp}(B_{1},B_{2}) \ =\ r_{\delta}^{(n+1)j-1}%
\ r_{\delta+j^{\prime}}\cdot\operatorname*{dp}(R_{\delta+j^{\prime}+e_{1}%
},\ldots,R_{\delta+j^{\prime}+e_{n}})
\end{align}
Combining \eqref{equ:003} and \eqref{equ:025}, we have
\begin{align*}
r_{\delta+j^{\prime}}^{n}\ r_{\delta}^{-n}\ R_{j^{\prime}+1^{\prime}%
}(R_{\delta+e_{0}},\ldots,R_{\delta+e_{n}}) \ =\ r_{\delta}^{(n+1)j-1}%
\ r_{\delta+j^{\prime}}\ {\operatorname*{dp}}(R_{\delta+j^{\prime}+e_{1}%
},\ldots, R_{\delta+j^{\prime}+e_{n}})
\end{align*}
which can be simplified into \eqref{equ:equivalent1} after pushing the
coefficients to the left-hand side. The proof of Eq.
\eqref{equ:theorem:k=j+1_1} is completed.
\end{proof}

\subsubsection{Proof of Eq. \eqref{equ:theorem:k=j+1_2} from Eqs.
\eqref{equ:theorem:k=j_1}--\eqref{equ:theorem:k=j_2}}

The goal of this part is to prove
\begin{align}
\label{equ:3.3.2need_to_proof}r_{w_{0}}^{(n+1)(j+1)}\ R_{w_{0}+j^{\prime
}+1^{\prime}+e_{i}} \ =\ R_{j^{\prime}+1^{\prime}+e_{i}}(R_{w_{0}+e_{0}%
},\ldots,R_{w_{0}+e_{n}}), \ \ i=1,\ldots,n
\end{align}
with \eqref{equ:theorem:k=j_1} and \eqref{equ:theorem:k=j_2} assumed to be true.

\begin{proof}
[Proof of Eq. \eqref{equ:theorem:k=j+1_2} ] Without loss of generality, we
proceed to prove it is true for the case when $i=1$. The other cases can be
proved with the same procedure.

By using Theorem \ref{thm:k=1} with $w_{0}$ specified as $w_{0}+j^{\prime}$,
we have
\begin{align*}
r_{w_{0}+j^{\prime}}^{n+1} R_{w_{0}+j^{\prime}+1^{\prime}+e_{1}}  &
\ =\ R_{1^{\prime}+e_{1}}(R_{w_{0}+j^{\prime}+e_{0}},\ldots,R_{w_{0}%
+j^{\prime}+e_{n}})\\
&  \ =\ {\operatorname*{dp}}(R_{w_{0}+j^{\prime}},xR_{w_{0}+j^{\prime}+e_{1}%
},R_{w_{0}+j^{\prime}+e_{1}},\ldots,R_{w_{0}+j^{\prime}+e_{n}})
\end{align*}
Let $\delta=w_{0}=(\delta_{1},\ldots,\delta_{n})$. Thus we only need to show
the following equivalence of \eqref{equ:3.3.2need_to_proof}:
\begin{align}
\label{equ:equivalent2} &  {\operatorname*{dp}}(R_{\delta+j^{\prime}%
},xR_{\delta+j^{\prime}+e_{1}},R_{\delta+j^{\prime}+e_{1}},\ldots
,R_{\delta+j^{\prime}+e_{n}}) \ =\ r_{\delta}^{-(n+1)(j+1)} r_{\delta
+j^{\prime}}^{n+1}\ R_{j^{\prime}+1^{\prime}+e_{1}}(R_{\delta+e_{0}}%
,\ldots,R_{\delta+e_{n}})
\end{align}
Let $\hat{G}_{i,s}=x^{s}R_{\delta+e_{i}}$ and $\hat{H}_{i,s}=(\hat{G}%
_{i,s},\ldots,\hat{G}_{i,0}$). Then we have
\begin{align*}
R_{j^{\prime}+1^{\prime}+e_{1}}(R_{\delta+e_{0}},\ldots,R_{\delta+e_{n}%
})=\operatorname*{dp}(\hat{H}_{0,j},\hat{H}_{1,j+1},\hat{H}_{2,j},\ldots
,\hat{H}_{n,j})
\end{align*}
Recall \eqref{equ:uij_1}, \eqref{equ:uij_2} and \eqref{equ:uij_3}. We further
deduce that
\begin{align}
\label{equ:026}r_{\delta}^{(n+1)j} R_{\delta+j^{\prime}+e_{i}} =\sum_{t=0}%
^{n}\sum_{h=0}^{j-1}u_{th}^{(i)}\hat{G}_{t,h}+u_{ij}^{(i)}\hat{G}_{i,j},
\ \ i=1\ldots,n
\end{align}
where $u_{ij}^{(i)}=(-1)^{(n+1-i)j} r_{\delta}^{(n+1)j-1} r_{\delta+j^{\prime
}}$. It follows that
\begin{align*}
u_{1j}^{(1)}\cdot\prod_{i=1}^{n}u_{ij}^{(i)} \ =\  &  (-1)^{nj} r_{\delta
}^{(n+1)j-1} r_{\delta+j^{\prime}} \cdot\left(  \prod_{i=1}^{n}(-1)^{(n+1-i)j}
r_{\delta}^{(n+1)j-1} r_{\delta+j^{\prime}}\right) \\
\ =\  &  (-1)^{nj}\left(  \prod_{i=1}^{n}(-1)^{(n+1-i)j}\right)
\cdot(r_{\delta}^{(n+1)j-1} r_{\delta+j^{\prime}})^{n+1}%
\end{align*}
Now we introduce the short-hands $\psi_{j} =(-1)^{nj}\prod_{i=1}%
^{n}(-1)^{(n+1-i)j}$ and $\rho_{j} = r_{\delta}^{(n+1)j}$. Then
\[
u_{1j}^{(1)}\cdot\prod_{i=1}^{n}u_{ij}^{(i)} \ =\ \psi_{j} \cdot(\rho_{j}
r_{\delta}^{-1} r_{\delta+j^{\prime}})^{n+1}%
\]
Next we consider the product of $\psi_{j}$, $(\rho_{j} r_{\delta}^{-1}
r_{\delta+j^{\prime}})^{n+1}$ and $R_{j^{\prime}+1^{\prime}+e_{1}}%
(R_{\delta+e_{0}},\ldots,R_{\delta+e_{n}})$ and carry out the following
simplification:
\begin{align*}
&  \psi_{j} \cdot(\rho_{j} r_{\delta}^{-1} r_{\delta+j^{\prime}})^{n+1}
R_{j^{\prime}+1^{\prime}+e_{1}}(R_{\delta+e_{0}},\ldots,R_{\delta+e_{n}})\\
\ =\  &  u_{1j}^{(1)}\prod_{i=1}^{n}u_{ij}^{(i)}\ \ \operatorname*{dp}(\hat
{H}_{0,j},\hat{H}_{1,j+1},\hat{H}_{2,j},\ldots,\hat{H}_{n,j})\\
\ =\  &  u_{1j}^{(1)}\prod_{i=1}^{n}u_{ij}^{(i)}\ \ \operatorname*{dp}(\hat
{H}_{0,j},\underset{\hat{H}_{1,j+1}}{\underbrace{\hat{G}_{1,j+1},\hat{G}%
_{1,j},\hat{H}_{1,j-1}}},\underset{\hat{H}_{2,j}}{\underbrace{\hat{G}%
_{2,j},\hat{H}_{2,j-1}}},\ldots,\underset{\hat{H}_{n,j}}{\underbrace{\hat
{G}_{n,j},\hat{H}_{n,j-1}}})\\
\ =\  &  \psi_{j}\cdot u_{1j}^{(1)}\prod_{i=1}^{n}u_{ij}^{(i)}%
\ \ \operatorname*{dp}(\hat{H}_{0,j},\hat{H}_{1,j-1},\hat{H}_{2,j-1}%
,\ldots,\hat{H}_{n,j-1},\hat{G}_{1,j+1},\hat{G}_{1,j},\hat{G}_{2,j}%
,\ldots,\hat{G}_{n,j})\\
&  \text{(by reordering the arguments of}\ {\operatorname*{dp}})\\
\ =\  &  \psi_{j}\ \operatorname*{dp}(\hat{H}_{0,j},\hat{H}_{1,j-1},\hat
{H}_{2,j-1},\ldots,\hat{H}_{n,j-1},u_{1j}^{(1)}\hat{G}_{1,j+1},u_{1j}%
^{(1)}\hat{G}_{1,j},u_{2j}^{(2)}\hat{G}_{2,j},\ldots,u_{nj}^{(n)}\hat{G}%
_{n,j})\\
&  \text{(by\ pushing}\ u_{ij}^{(i)}\ \text{into} \ {\operatorname*{dp}})\\
\ =\  &  \psi_{j}\ {\operatorname*{dp}}(\hat{H}_{0,j},\hat{H}_{1,j-1},\hat
{H}_{2,j-1},\ldots,\hat{H}_{n,j-1},\rho_{j}\ xR_{\delta+j^{\prime}+e_{1}}%
,\rho_{j}\ R_{\delta+j^{\prime}+e_{1}},\rho_{j}\ R_{\delta+j^{\prime}+e_{2}%
},\ldots,\rho_{j}\ R_{\delta+j^{\prime}+e_{n}})\\
&  \text{(by}\ \eqref{equ:026})\\
\ =\  &  \psi_{j}\ \rho_{j}^{n+1}\ {\operatorname*{dp}}(\hat{H}_{0,j},\hat
{H}_{1,j-1},\hat{H}_{2,j-1},\ldots,\hat{H}_{n,j-1},xR_{\delta+j^{\prime}%
+e_{1}},R_{\delta+j^{\prime}+e_{1}}, R_{\delta+j^{\prime}+e_{2}},\ldots,
R_{\delta+j^{\prime}+e_{n}})\\
&  \text{(by\ pulling\ out }\rho_{j}\ \text{out\ of }{\operatorname*{dp}})
\end{align*}
The cancellation of $\psi_{j}\rho_{j}^{n+1}$ from the first and last
expressions yields
\begin{align}
&  r_{\delta}^{-n-1}r_{\delta+j^{\prime}}^{n+1}R_{j^{\prime}+1^{\prime}+e_{1}%
}(R_{\delta+e_{0}},\ldots,R_{\delta+e_{n}})\nonumber\\
\ =\  &  {\operatorname*{dp}}(\hat{H}_{0,j},\hat{H}_{1,j-1},\hat{H}%
_{2,j-1},\ldots,\hat{H}_{n,j-1},xR_{\delta+j^{\prime}+e_{1}},R_{\delta
+j^{\prime}+e_{1}}, R_{\delta+j^{\prime}+e_{2}},\ldots,R_{\delta+j^{\prime
}+e_{n}}) \label{eq:partitioned_matrix2}%
\end{align}
Partitioning the polynomial matrix in the right-hand side of
\eqref{eq:partitioned_matrix2}, we have
\begin{align}
\label{equ:007}r_{\delta}^{-n-1}r_{\delta+j^{\prime}}^{n+1}R_{j^{\prime
}+1^{\prime}+e_{1}}(R_{\delta+e_{0}},\ldots,R_{\delta+e_{n}}%
)={\operatorname*{dp}}(B_{1},B_{2})
\end{align}
where
\begin{align*}
B_{1}=(\hat{H}_{0,j},\hat{H}_{1,j-1},\hat{H}_{2,j-1},\ldots,\hat{H}%
_{n,j-1}),\ \ \ B_{2}=(xR_{\delta+j^{\prime}+e_{1}},R_{\delta+j^{\prime}%
+e_{1}}, R_{\delta+j^{\prime}+e_{2}},\ldots, R_{\delta+j^{\prime}+e_{n}})
\end{align*}
Denote $\max(\deg B_{1})$, $\max(\deg B_{2})$ and the number of polynomials in
$B_{1}$ by $m_{1}$, $m_{2}$ and $\#B_{1}$, respectively. Then we have
$m_{1}=d_{0}-|\delta|+j$, $m_{2}=d_{0}-|\delta|-nj$ and $\#B_{1}=(n+1)j+1$. It
follows that
\begin{align*}
m_{2}=m_{1}-\#B_{1}+1
\end{align*}
By Lemma \ref{lemma:002}-(1), we derive the following
\begin{align}
\label{equ:028}\operatorname*{dp}(B_{1},B_{2}) =\operatorname*{dp}%
({\operatorname*{dp}}(B_{1}),B_{2})
\end{align}
In order to have an explicit expression of $\operatorname*{dp}(B_{1})$, we
carry out the following deduction:
\begin{align}
\label{equ:029}\operatorname*{dp}(B_{1}) =  &  \operatorname*{dp}(\hat
{H}_{0,j},\hat{H}_{1,j-1},\hat{H}_{2,j-1},\ldots,\hat{H}_{n,j-1})\nonumber\\
=  &  \operatorname*{dp}(\underset{\hat{H}_{0,j}}{\underbrace{\hat{G}%
_{0,j},\hat{G}_{0,j-1},\hat{H}_{0,j-2}}},\hat{H}_{1,j-1},\hat{H}%
_{2,j-1},\ldots,\hat{H}_{n,j-1})\nonumber\\
=  &  {\operatorname*{pcdp}}(\hat{G}_{0,j},\hat{G}_{0,j-1})\cdot
\operatorname*{dp}(\hat{H}_{0,j-2},\hat{H}_{1,j-1},\hat{H}_{2,j-1},\ldots
,\hat{H}_{n,j-1})\ \ \ \ \ \text{by\ Lemma\ \ref{lemma:002}-(2)}\nonumber\\
=  &  r_{\delta}^{2}\cdot R_{j^{\prime}}(R_{\delta},\ldots,R_{\delta^{\prime
}+e_{n}})\nonumber\\
=  &  r_{\delta}^{2}\cdot r_{\delta}^{(n+1)j-2} R_{\delta+j^{\prime}%
}\ \ \ \ \ \text{by\ the\ assumption}\ \eqref{equ:theorem:k=j_1}\nonumber\\
=  &  r_{\delta}^{(n+1)j}\ R_{\delta+j^{\prime}}%
\end{align}
After substituting \eqref{equ:029} into \eqref{equ:028}, we have
\begin{align}
\operatorname*{dp}(B_{1},B_{2})  &  = \operatorname*{dp}\ ( r_{\delta
}^{(n+1)j} R_{\delta+j^{\prime}},\ \ \ xR_{\delta+j^{\prime}+e_{1}},
R_{\delta+j^{\prime}+e_{1}},R_{\delta+j^{\prime}+e_{2}},\ldots,R_{\delta
+j^{\prime}+e_{n}})\nonumber\\
&  = r_{\delta}^{(n+1)j}\ {\operatorname*{dp}}\ (\ R_{\delta+j^{\prime}%
},\ \ xR_{\delta+j^{\prime}+e_{1}}, R_{\delta+j^{\prime}+e_{1}},R_{\delta
+j^{\prime}+e_{2}},\ldots,R_{\delta+j^{\prime}+e_{n}}) \label{equ:008}%
\end{align}
Combining \eqref{equ:007} and \eqref{equ:008}, we have
\begin{align*}
&  r_{\delta}^{-n-1}r_{\delta+j^{\prime}}^{n+1}R_{j^{\prime}+1^{\prime}+e_{1}%
}(R_{\delta+e_{0}},\ldots,R_{\delta+e_{n}})\\
=  &  r_{\delta}^{(n+1)j}{\operatorname*{dp}}(R_{\delta+j^{\prime}}%
,xR_{\delta+j^{\prime}+e_{1}}, R_{\delta+j^{\prime}+e_{1}},R_{\delta
+j^{\prime}+e_{2}},\ldots,R_{\delta+j^{\prime}+e_{n}})
\end{align*}
which can be simplified into \eqref{equ:equivalent2} after pushing the
coefficients to the left-hand side. The proof of Eq.
\eqref{equ:theorem:k=j+1_2} is completed.
\end{proof}

\section{Conclusion}

\label{sec:conclusion} In this paper, we present a generalization of the well
known Habicht's theorem to several polynomials, that is, expressing a
subresultant of several polynomials with higher order to the subresultant of
other subresultants with lower orders. With the help of this discovery, we can
identify many non-trivial inherent relationships among subresultants of
several polynomials and thus can explore the nice hidden structure of
subresultants in a more efficient and systematic way.

In the classical subresultant theory, Habicht's theorem has been proved to be
a productive tool for the theory development, from which people discovered the
famous gap structure of subresultant chain for two polynomials. We hope that
the generalized Habicht's theorem can also become a powerful tool for people
to investigate similar structures for subresultants of multiple polynomials.

One may also notice that in the classical Habicht's theorem, in some special
degeneracy cases, the subresultant of two polynomials is proportional to the
pseudo-remainder of other two subresultants, which means the Habicht's theorem
can be stated in a uniform way. However, for the multi-polynomial case, it is
no longer true. That is, the result in \cite{Hong_Yang-2023} and that in the
current paper are independent. Therefore, a natural question is how to
formulate them in a uniform way, which is also worthy of further investigation
in the future.

\bigskip\noindent\textbf{Acknowledgements.} Hoon Hong's work was supported by
National Science Foundations of USA (Grant Nos: CCF 2212461 and CCF 2331401).
Jing Yang and Jiaiqi Meng's work was supported by National Natural Science Foundation of China
(Grant Nos.: 12261010 and 12326353) and the Natural Science Cultivation
Project of GXMZU (Grant No.: 2022MDKJ001).


\def\cprime{$'$}

\end{document}